\documentclass[aps,showpacs,nofootinbib,superscriptaddress,notitlepage]{revtex4-1}
\usepackage{epsf}
\usepackage{graphicx}
\usepackage{amsmath}
\usepackage{mathrsfs}
\usepackage{hyperref}
\usepackage{microtype}

\def\slashchar#1{\setbox0=\hbox{$#1$}
   \dimen0=\wd0 \setbox1=\hbox{/} \dimen1=\wd1
   \ifdim\dimen0>\dimen1 \rlap{\hbox to \dimen0{\hfil/\hfil}} #1
   \else  \rlap{\hbox to \dimen1{\hfil$#1$\hfil}} / \fi}

\newcommand{\identidad}{\mbox{1}\hspace{-0.35ex}\rule{0.1mm}{1.5ex}\hspace{1ex}}
\newcommand{\reales}{\mbox{R}\hspace{-1.0ex}\rule{0.1mm}{1.5ex}\hspace{1ex}}
\begin{document}

\title{Watson's theorem  and the $N\Delta(1232)$ axial transition} 

\author{L.~Alvarez-Ruso}
\affiliation{Instituto de F\'\i sica Corpuscular (IFIC), Centro Mixto
CSIC-Universidad de Valencia, Institutos de Investigaci\'on de
Paterna, E-46071 Valencia, Spain} 
\author{E. Hern\'andez} \affiliation{Departamento de F\'\i sica Fundamental 
e IUFFyM,\\ Universidad de Salamanca, E-37008 Salamanca, Spain} 
\author{J.~Nieves}
\affiliation{Instituto de F\'\i sica Corpuscular (IFIC), Centro Mixto
CSIC-Universidad de Valencia, Institutos de Investigaci\'on de
Paterna, E-46071 Valencia, Spain} 
\author{M.J. \surname{Vicente Vacas}}
\affiliation{Departamento de F\'\i sica Te\'orica and IFIC, Centro Mixto
Universidad de Valencia-CSIC, Institutos de Investigaci\'on de
Paterna, E-46071 Valencia, Spain}


\begin{abstract}
We present a new determination of the $N\Delta$ axial form factors from neutrino 
induced pion production data. For this purpose, the model of Hernandez 
{\it et al.}, Phys. Rev. D 76, 033005 (2007) is improved by partially restoring unitarity. This is accomplished by imposing
 Watson's theorem on the dominant vector and axial multipoles. As a consequence,
 a larger $C_5^A(0)$, in  good agreement with the prediction from the off-diagonal Goldberger-Treiman relation, is now obtained.
\end{abstract}

\maketitle

\section{Introduction}

Weak pion production off nucleons provides  valuable insight into the axial 
structure of hadrons. In addition, pion production cross sections grow to 
become one of the main reaction mechanisms for neutrinos of few-GeV energies, 
which is an important range for current and future oscillation experiments. 
Therefore, a better understanding of weak pion production mechanisms is 
actively pursued~\cite{Morfin:2012kn,Formaggio:2013kya,Alvarez-Ruso:2014bla}. 
Recent measurements on, predominantly, carbon targets by 
MiniBooNE~\cite{AguilarArevalo:2009ww,AguilarArevalo:2010xt,
AguilarArevalo:2010bm} and MINERvA~\cite{AguilarArevalo:2010bm,Aliaga:2015wva} 
experiments have revealed discrepancies with existing theoretical models and 
among different data sets~\cite{Lalakulich:2012cj,Hernandez:2013jka,
Yu:2014yja,Sobczyk:2014xza,Mosel:2015tja}.

The first requirement to achieve a precise knowledge of neutrino
 induced pion production on nuclear targets is a realistic model at the
  nucleon level. Theoretical studies of weak pion production off the nucleon 
  at intermediate energies~\cite{Adler:1968tw,Bijtebier:1970ku,Zucker:1971hp,
  LlewellynSmith:1971zm,Schreiner:1973mj,Alevizos:1977xf,Fogli:1979cz,
  Fogli:1979qj,Rein:1980wg,Hemmert:1994ky,AlvarezRuso:1997jr,
  AlvarezRuso:1998hi,Slaughter:1998cw,Golli:2002wy,Sato:2003rq,
  Paschos:2003qr,Lalakulich:2005cs,Lalakulich:2006sw,Hernandez:2007qq,
  Graczyk:2007bc,Leitner:2008ue,Barbero:2008zza,Graczyk:2009qm,
  Hernandez:2010bx,Lalakulich:2010ss,Serot:2012rd,
  Barbero:2013eqa,Zmuda:2015rea,Nakamura:2015rta,Alam:2015gaa} have 
  highlighted the important role of baryon resonance excitation, 
  predominantly the $\Delta(1232)3/2^+$. The weak nucleon-to-$\Delta(1232)$ 
  transition current can be written in terms of vector and axial form 
  factors, $C_{3-5}^V$ and $C_{3-6}^A$ in the notation of 
  Ref.~\cite{LlewellynSmith:1971zm}. Although there are quark model 
  determinations of these form factors~\cite{Feynman:1971wr,Liu:1995bu,
  Golli:2002wy,BarquillaCano:2007yk}, a common strategy is to adopt 
  empirical parametrizations for them. The role of heavier resonances 
  has also been investigated although the available experimental 
  information about the axial sector is very limited. Among these states, 
  only the $N(1520)3/2^-$ appears to be relevant for neutrino energies below
   1.5~GeV~\cite{Leitner:2008ue}. Nonresonant electroweak amplitudes have 
   also been extensively considered.  As pointed out in 
   Ref.~\cite{Hernandez:2007qq}, these terms are not only demanded but, 
   close to threshold, fully fixed by chiral symmetry. Away from threshold, 
   these amplitudes are usually modeled using phenomenologically parametrized 
   nucleon form factors, introduced in a way that respects both the
    conservation of the vector current (CVC) and the partial conservation 
    of the axial current (PCAC). 

In Ref.~\cite{Hernandez:2007qq} (referred from now on as the HNV model)
 nonresonant amplitudes, evaluated from the leading contributions of
  the SU(2) chiral Lagrangian, supplemented with empirical parametrizations
   of the nucleon form factors, were considered alongside  the 
   $\Delta(1232)$ excitation. The vector form factors in the $N\Delta$ 
   vertex come from helicity amplitudes extracted in the analysis of 
   electron scattering data~\cite{Lalakulich:2005cs}. The most important 
   among the axial form factors is $C^A_5$, which appears at leading 
   order in an expansion of the hadronic tensor in the four-momentum 
   transfer $q^2$. Assuming the pion pole dominance of the pseudoscalar 
   form factor $C^A_6$, it can be related to $C^A_5$ owing to PCAC. For 
   the  subleading $C_{3,4}^A$ form factors,  Adler's 
   parametrizations~\cite{Adler:1968tw,Bijtebier:1970ku} were 
   adopted: $C_3^A=0, C_4^A=-C_5^A/4$. The available bubble-chamber 
   data on pion production induced by neutrinos on deuterium, taken at
    Argonne and Brookhaven National Laboratories 
    (ANL and BNL)~\cite{Radecky:1981fn,Kitagaki:1986ct} are quite 
    insensitive to the values of these form factors~\cite{Hernandez:2010bx}. 
    With the aim of extending the model toward higher energies, the $N(1520)$
     intermediate state was added in Ref.~\cite{Hernandez:2013jka} using the
      transition form factors introduced in Ref.~\cite{Leitner:2008ue}.

The pion pole dominance of $C^A_6$ and PCAC result in a relation between 
the leading axial coupling $C_5^A(q^2=0)$ and the $\Delta \rightarrow N \pi$ 
decay coupling known as the off-diagonal Goldberger-Treiman relation (GTR). Studies
 that neglected the nonresonant contributions found  good agreement between 
 the $C_5^A(0)$ value extracted from ANL and/or BNL data and the
  GTR~\cite{AlvarezRuso:1998hi,Graczyk:2009qm}. However, the fit 
  of $C_5^A(q^2)$ to the flux averaged $\nu_\mu p \to \mu^- p\pi^+$ 
  ANL  $q^2$-differential cross section data~\cite{Radecky:1981fn} 
  with the HNV model found a discrepancy of 30\% with respect to the GTR 
  prediction of $C_5^A(0) = 1.15-1.2$. A simultaneous fit to both ANL and 
  BNL data samples including independent overall flux normalization 
  uncertainties for each experiment, as suggested in Ref.~\cite{Graczyk:2009qm}, 
  and considering deuterium-target corrections obtained 
  $C_5^A(0) = 1.00\pm 0.11$~\cite{Hernandez:2010bx}, still 2$\sigma$ below 
  the GTR value. Although the HVN model could be reconciled with the GTR by 
  simultaneously fitting vector form factors to electron-proton scattering 
  structure function $F_2$~\cite{Graczyk:2014dpa,Zmuda:2015rea}, it should be 
  realized that the HNV model does not satisfy  Watson's 
  theorem~\cite{Watson:1952ji}. The latter, which is a consequence of 
  unitarity and time-reversal invariance, implies that the phase of the 
  electroweak pion production is fully determined by the strong $\pi N$ 
  interaction. The goal of the present study is to impose  Watson's 
  theorem in the HNV model.  It is shown that, in this way, the consistency 
  with the GTR prediction is restored.

 The dynamical model of photo-, electro- and weak
pion production  derived in Ref.~\cite{Sato:2003rq} deserves a special mention. 
To date, this is the
only weak pion production model fulfilling Watson's theorem exactly. Starting 
from an
effective Hamiltonian with bare $N\Delta$ couplings obtained
in a nonrelativistic constituent quark model~\cite{Hemmert:1994ky}, the
Lippmann-Schwinger equation in coupled channels is solved, which restores
unitarity. Besides, the bare couplings get renormalized by meson clouds.
The predicted cross sections are in good agreement with
data (Figs. 5-8 of Ref.~\cite{Sato:2003rq}). The scheme has been further
refined and extended to incorporate $N^*$ resonances and a larger number of 
meson-baryon states~\cite{Kamano:2012id,Nakamura:2015rta}. Although the 
chiral counting at threshold is broken by the presence of $\rho$ and $\omega$ 
exchanges in the $t$-channel or the introduction of explicit $\sigma$ meson 
intermediate states, this framework should satisfy unitary constraints and 
fulfill Watson's theorem. The partially unitarized HNV model presented here
 is considerably simpler. The agreement with the GTR and a good description 
 of data for invariant masses $W_{\pi N} < 1.4$~GeV are achieved by introducing 
 two relative phases between the $\Delta(1232)$ and the nonresonant 
contributions. The HNV model improved in this way is portable and can 
be easily implemented in event generators used in the analysis of neutrino 
oscillation experiments.

The paper is organized as follows. In Sec.~\ref{sec:unitarity}, we introduce
Watson's theorem, which is based on unitarity and time reversal invariance, 
and explain its implementation in the HNV model. In Sec.~\ref{sec:results} we
present the new extraction of the $C_5^A(q^2)$ axial form factor. 
Appendices~\ref{sec:states},\ref{app:chi} and \ref{app:chiHNV} collect some
 useful formulas
needed for the calculation. Finally, in Appendix~\ref{app:fits} we give a
parametrization of the Olsson phases (see below) used to impose Watson's
theorem in our approach. 
 
\section{Unitarity, time-reversal invariance and Watson's theorem}
\label{sec:unitarity}

A scattering process due to short-range interactions (like strong or weak 
interactions) can be described in terms of initial and final states of 
noninteracting particles. The amplitude for a
transition is given by the corresponding matrix element of the scattering 
operator
\begin{equation}
S=\identidad - i T \,.
\end{equation}

Given an initial state $| I\rangle$, the probability for finding the system 
in an asymptotic state
$| N \rangle$ is $P_N=|\langle N | S | I\rangle|^2$; since $\sum_NP_N=1$,
 one deduces that $S$ is a unitary
operator, $S S^\dagger = S^\dagger S =\identidad$, which implies 
that\footnote{The optical theorem trivially follows from the particular case
$|I\rangle = |F \rangle$,
\begin{equation}
{\rm Im}\langle I | T | I \rangle = -\frac12 \sum_N |\langle N|T| I\rangle|^2.
\end{equation}}  
\begin{eqnarray}
i(T-T^\dagger) &=& T^\dagger T \, \nonumber\\
i\big \{\langle F | T | I
\rangle -\langle F | T^\dagger | I
\rangle\big \}&=& \langle F| T^\dagger T| I\rangle = \sum_N \langle F
| T^\dagger|N\rangle  \langle N|T| I\rangle = \sum_N \langle N
|T|F\rangle^*  \langle N|T| I\rangle  \,.\label{eq:unitarity}
\end{eqnarray}
On the other hand, if time reversal invariance holds,
\begin{equation}
\langle F| S | I\rangle = \langle I_{\mathscr{T}} | S | F_{\mathscr{T}} \rangle \,,
\end{equation}
where $\mathscr{T} |I\rangle = |I_{\mathscr{T}} \rangle$ and 
$\mathscr{T} |F\rangle = |F_{\mathscr{T}} \rangle$. In other words, if the 
system is time reversal invariant, 
$\mathscr{T} S \mathscr{T}^\dagger = S^\dagger$ and therefore
$\mathscr{T}^\dagger T^\dagger \mathscr{T}   = T$. The time reversal operator $\mathscr{T}$ is antiunitary\footnote{This is to say antilinear, $\langle
  A_{\mathscr{T}} |O| B_{\mathscr{T}}  \rangle =\langle
  A|\mathscr{T}^\dagger O\mathscr{T} |B \rangle^* $, and satisfying
  $\mathscr{T}^{-1}= \mathscr{T}^\dagger$.}~\cite{martin1970elementary,gibson1980symmetry} with
$\mathscr{T}^2=\pm \identidad$. Thus,  one finds
\begin{equation}
\langle F | T^\dagger | I\rangle = \langle I | T | F\rangle^*  =
\langle I | \mathscr{T}^\dagger T^\dagger \mathscr{T}  | F\rangle^* = 
 \langle I_{\mathscr{T}} | T^\dagger   | F_{\mathscr{T}}\rangle =
 \langle F_{\mathscr{T}} | T   | I_{\mathscr{T}}\rangle^* \,.
\end{equation}
Using this result in Eq.~(\ref{eq:unitarity}), we obtain from unitarity
and time reversal invariance that 
\begin{equation}
i\big \{\langle F | T | I
\rangle -\langle F_{\mathscr{T}} | T   | I_{\mathscr{T}}\rangle^* \big
\} = \sum_N \langle N
|T|F\rangle^*  \langle N|T| I\rangle \label{eq:watson-prel}
\end{equation}
If $\langle F | T | I \rangle = \langle F_{\mathscr{T}} | T   |
I_{\mathscr{T}}\rangle $, which is always satisfied 
for transitions between center of mass (CM) two-particle states with well defined
helicities and total angular momentum  whenever the 
interaction is invariant under time
reversal~\cite{martin1970elementary}, and there is only one relevant
intermediate state in the sum of Eq.~(\ref{eq:watson-prel}), one obtains that
\begin{equation}
\langle N |T|F\rangle^*  \langle N|T| I\rangle = -2\,{\rm Im}\langle F | T | I
\rangle \in \reales \,
\end{equation}
so that the phases of $\langle N |T|F\rangle$ and $\langle N|T|
I\rangle$ coincide. This result constitutes Watson's 
theorem~\cite{Watson:1952ji} on the effect of final state interactions
on reaction cross sections. As shown, it is a consequence of
unitarity and time reversal invariance.

\subsection{Watson's theorem for CM two-particle helicity states}

Assuming that only two-particle intermediate states (2body), with
masses $m_1^{\prime}$ and $m_2^{\prime}$,  contribute\footnote{This is exact below
the three-particle threshold.}, the unitarity condition of
Eq.~(\ref{eq:unitarity}) for the binary process 
$a+b\to 1+2$ can be written as 
\begin{eqnarray}
&&i\left\{\langle
\theta,\varphi;\lambda_1,\lambda_2;\gamma_{12}\,|T(s)|0,0;\lambda_a,\lambda_b;\gamma_{ab}\,\rangle-
\langle
\theta,\varphi;\lambda_1,\lambda_2;\gamma_{12}\,|T^\dagger(s)|0,0;\lambda_a,\lambda_b;\gamma_{ab}\,\rangle\right\}
=\sum_{\rm 2body}\frac{\lambda^{1/2}(s,m_1^{\prime\,2},m_2^{\prime\,2})}{32\pi^2s}
\nonumber \\
&\times& \int d\Omega' \sum_{\gamma^{\,\prime}_{12}}\sum_{\lambda'_1\lambda'_2}
\langle\theta',\varphi';\lambda'_1,\lambda'_2;\gamma^{\,\prime}_{12}\,|T(s)|\theta,\varphi;\lambda_1,\lambda_2;\gamma_{12}\,\rangle^*
\langle\theta',\varphi';\lambda'_1,\lambda'_2;\gamma^{\,\prime}_{12}\,|T(s)
|0,0;\lambda_a,\lambda_b;\gamma_{ab}\,\rangle \, \label{eq:watson1}
\end{eqnarray}
where $s=(p_a+p_b)^2$ and the function $\lambda(x,y,z)=x^2+y^2+z^2-2 x y -2 x z -2 y z$. Two-particle states in the CM are defined in Appendix~\ref{sec:states}. The matrix element of the $T$ operator is computed in
the little Hilbert space (see Appendix~\ref{sec:states} and the book of Martin \& Spearman~\cite{martin1970elementary})
\begin{equation}
\langle
\theta,\varphi;\lambda_1,\lambda_2;\gamma_{12}\,|T(s)|0,0;\lambda_a,\lambda_b;\gamma_{ab}\,\rangle
\equiv \langle \alpha_F | T_P | \alpha_I \rangle = \frac{(2\pi)^2
  4\sqrt{s}}{\sqrt{|\vec{p}_I||\vec{p}_F|}}
\langle\theta,\varphi;\lambda_1,\lambda_2;\gamma_{12}\,|T_P|0,0;\lambda_a,\lambda_b;\gamma_{ab}\,\rangle\equiv T_{FI}(s) \label{Tp}
\end{equation}
where $|\vec{p}_{I(F)}| \equiv |\vec{p}_{a(1)}| = |\vec{p}_{b(2)}|$; $T_P$ is the reduction 
of the full operator $T$ in the little Hilbert space, Eq.~(\ref{eq:tp}). 
Normalizations are fixed by the expression of the CM differential cross 
section for the $a+b\to 1+2$ reaction, which is calculated as 
\begin{equation}
\frac{d\sigma}{d\Omega} = \frac{1}{64\pi^2s}\frac{|\vec{p}_F|}{|\vec{p}_I|}\left|T_{FI}(s)\right|^2
\end{equation}
The unitarity condition, Eq.~(\ref{eq:watson1}), can be rewritten for states
 $|J,M\rangle$ with 
well-defined angular momentum. Changing basis with Eq.~(\ref{eq:cJ}) and 
using the orthogonality properties of the 
${\cal D}^{(J)}_{MM'}(\varphi,\theta,-\varphi)$ rotation 
matrices [Eq.~(\ref{eq:ortD})], the condition 
${\cal D}^{(J)}_{MM'}(0,0,0) =\delta_{MM'}$, the fact that 
$T$ is a scalar under rotations, and  Parseval's identity 
associated to Eq.~(\ref{eq:ortJM}), one gets that
\begin{eqnarray}
&i&\left\{\langle J,M;\lambda_1,\lambda_2;\gamma_{12}\,|T(s)|J,M;\lambda_a,\lambda_b;\gamma_{ab}\,\rangle-
\langle
J,M;\lambda_1,\lambda_2;\gamma_{12}\,|T^\dagger(s)|J,M;\lambda_a,\lambda_b;\gamma_{ab}\,\rangle\right\} =
\sum_{\rm 2body}\frac{\lambda^{1/2}(s,m_1^{\prime\,2},m_2^{\prime\,2})}{32\pi^2s}
\nonumber \\
&\times& \sum_{\gamma^{\,\prime}_{12}}\sum_{\lambda'_1\lambda'_2} \langle J,M;\lambda'_1,\lambda'_2;\gamma^{\,\prime}_{12}\,|T(s)|J,M;\lambda_1,\lambda_2;\gamma_{12}\,
\rangle^* \langle J,M;\lambda'_1,\lambda'_2;\gamma^{\,\prime}_{12}\,|T(s)|J,M;\lambda_a,\lambda_b;\gamma_{ab}\,\rangle\,,
\label{eq:watson2}
\end{eqnarray}
with $M=\lambda_a-\lambda_b$ as follows from Eq.~(\ref{eq:whyMisfixed}). In practice, all the above matrix
elements do not  depend on $M$ because $T$ is a scalar under rotations. Hence, it is usual to adopt the short notation 
\begin{equation}
\langle
J,M;\lambda_1,\lambda_2;\gamma_{12}\,|T(s)|J,M;\lambda_a,\lambda_b;\gamma_{ab}\,\rangle
= \langle \lambda_1,\lambda_2;\gamma_{12}\,|T_J(s)|\lambda_a,\lambda_b;\gamma_{ab}\,\rangle
\end{equation}
Assuming time reversal invariance ($\mathscr{T}^\dagger
T\mathscr{T}=T^\dagger$), 
\begin{equation}
\langle
\lambda_1,\lambda_2;\gamma_{12}\,|T_J^\dagger(s)|\lambda_a,\lambda_b;\gamma_{ab}\,\rangle
=\langle
\lambda_1,\lambda_2;\gamma_{12}\,|\mathscr{T}^\dagger T_J(s)\mathscr{T}|\lambda_a,\lambda_b;\gamma_{ab}\,\rangle
= \langle
\lambda_1,\lambda_2;\gamma_{12}\,|T_J(s)|\lambda_a,\lambda_b;\gamma_{ab}\,\rangle^*
\label{eq:1glup}
\end{equation}
since $\mathscr{T}$ is an antiunitary operator. We have also used the 
transformation properties under time reversal 
of the helicity $|J,M;\lambda,\lambda';\gamma\,\rangle$
states\footnote{To obtain Eq.~(\ref{eq:2glup}), the intrinsic time reversal 
parities of all involved particles have been set to +1 (see 
Ref.~\cite{martin1970elementary}). Within the conventions used in
Ref.~\cite{Hernandez:2007qq} (HNV model), this is not the case for the
pion, which should be taken into account in the following (see the 
discussion in Appendix~\ref{app:chiHNV}).}
\begin{eqnarray}
\mathscr{T}
|J,M;\lambda_1,\lambda_2;\gamma\,\rangle=(-1)^{J-M}|J,-M;\lambda_1,\lambda_2;\gamma\,\rangle,\label{eq:2glup}
\end{eqnarray}
Thus, the left-hand side of  Eq.~(\ref{eq:watson2}) becomes 
\begin{equation}
i\left\{\langle \lambda_1,\lambda_2;\gamma_{12}\,|T_J(s)|\lambda_a,\lambda_b;\gamma_{ab}\,\rangle-
\langle \lambda_1,\lambda_2;\gamma_{12}\,|
T_J^\dagger(s)|\lambda_a,\lambda_b;\gamma_{ab}\,\rangle\right\}=-2 {\rm
  Im}\langle
\lambda_1,\lambda_2;\gamma_{12}\,|T_J(s)|\lambda_a,\lambda_b;\gamma_{ab}\,\rangle \label{eq:4glup}
\end{equation}
Hence (provided time reversal invariance holds) one finds
\begin{eqnarray}
{\rm
  Im}\langle
\lambda_1,\lambda_2;\gamma_{12}\,|T_J(s)|\lambda_a,\lambda_b;\gamma_{ab}\,\rangle
&=& -\frac12 \sum_{\rm
  2body}\frac{\lambda^{1/2}(s,m_1^{\prime\,2},m_2^{\prime\,2})}{32\pi^2s} \nonumber \\ 
& \times &
\sum_{\gamma^{\,\prime}_{12}}\sum_{\lambda'_1\lambda'_2}
\langle \lambda'_1,\lambda'_2;\gamma^{\,\prime}_{12}\,|T_J(s)|\lambda_1,\lambda_2;\gamma_{12}\,
\rangle^*
\langle\lambda'_1,\lambda'_2;\gamma^{\,\prime}_{12}\,|T_J(s)|\lambda_a,\lambda_b;\gamma_{ab}\,\rangle  \label{eq:watson3}
\end{eqnarray}
Let us consider an electroweak transition from an initial state
($a+b$) involving at least a gauge boson, to a purely hadronic final state ($1+2$). Furthermore,
let us assume that the total c.m. energy, $\sqrt{s}$, is such that
the only relevant strong process is the elastic one $1+2\to 1+2$.
In these circumstances, the sum over intermediate states in Eq.~(\ref{eq:watson3}) is 
dominated by the  $1+2\to 1+2$ strong $T$ matrix. The contribution of any other
intermediate state will be proportional to the product of two electroweak
transition amplitudes, and hence highly suppressed. Therefore, 
\begin{equation}
\sum_{\gamma^{\,\prime}_{12}}\sum_{\lambda'_1\lambda'_2}
\langle \lambda'_1,\lambda'_2;\gamma^{\,\prime}_{12}\,|T_J(s)|\lambda_1,\lambda_2;\gamma_{12}\,
\rangle^*
\langle\lambda'_1,\lambda'_2;\gamma^{\,\prime}_{12}\,|T_J(s)|\lambda_a,\lambda_b;\gamma_{ab}\,\rangle
\in \reales \,, \label{eq:watJM}
\end{equation}
which establishes a series of relations between the phases of the
electroweak $a+b\to 1+2$ and the strong $1+2\to 1+2$ amplitudes.
\subsection{Watson's theorem for $WN\to\pi N$ and $ZN\to\pi N$ amplitudes}
\label{sec:wtwz}
Pion production off nucleons induced by (anti)neutrinos proceeds through charged (CC) or
neutral current (NC) interactions. These are determined by transition amplitudes of the
kind $WN\to\pi N$ and $ZN\to\pi N$, respectively.  In the following, we explicitly refer to the
CC case, but the extension to NC processes is straightforward. The off-shell-ness of the $W$ boson does not alter the following arguments and will be reconsidered later on.

For the $W N\to \pi N$ reaction, considering only $\pi N$ intermediate states, Eq.~(\ref{eq:watJM}) becomes\footnote{As the states are fully defined, the sum over $\gamma'_{12}$ can be dropped.}
\begin{eqnarray}
\sum_{\rho}
\langle J,M;\underbrace{0,\rho}_{\pi N}|T(s)|J,M;\underbrace{0,\lambda'}_{\pi N}
\rangle^*\langle J,M;\underbrace{0,\rho}_{\pi N}|T(s)|J,M;\underbrace{r,\lambda}_{W
N}\rangle
\in \reales,\qquad M=r-\lambda
\end{eqnarray}
where $r$ is the  helicity of the $W$ gauge boson  and
$\lambda, \lambda', \rho$ are the corresponding helicities of the initial,
final and intermediate nucleons.  The above expression is equivalent
to\footnote{We use that
  $|0,0;r,\lambda\,\rangle=\sum_J\sqrt{\frac{2J+1}{4\pi}}\,
  |J, M=r-\lambda;r,\lambda\,\rangle$
 and that $T$ is a scalar.}
\begin{eqnarray}
\sum_{\rho}
\langle J,M;\underbrace{0,\rho}_{\pi N}|T(s)|J,M;\underbrace{0,\lambda'}_{\pi N}
\rangle^*\langle J,M;\underbrace{0,\rho}_{\pi N}|T(s)|0,0;\underbrace{r,\lambda}_{W
N}\rangle
\in\reales.
\end{eqnarray}
where, we identify the initial $W N$ pair with the $z$ direction ($\theta=0,\varphi=0$) helicity CM
two-particle state.  Introducing states with well-defined orbital angular momentum $L$ and spin $S$ [Eq.~(\ref{eq:LSstates})], 
and using parity conservation on the $\pi N\to\pi N$ matrix elements, one gets 
\begin{eqnarray}
\sum_{L}\sum_{\rho}\frac{2L+1}{2J+1} (L,1/2,J|0,-\lambda',-\lambda')
(L,1/2,J|0,-\rho,-\rho)&&
\nonumber\\
\times\underbrace{\langle J,M;{L,1/2}|T(s)|J,M;{L,1/2}
\rangle^*}_{\pi N\to \pi N}\underbrace{\langle
  J,M;{0,\rho}|T(s)|0,0;{r,\lambda}\rangle}_{WN\to \pi N}
&\in\reales&\qquad \forall J,\, M=r-\lambda. \label{eq:watson-gral}
\end{eqnarray}
Here $(L,S,J|M_L,M_S,M_J)$ are Clebsch-Gordan coefficients. 

\subsection{Olsson's implementation of Watson's theorem for the $WN\to\pi N$
  amplitude in the $\Delta$ region. }
\label{sec:olss}

At intermediate energies, the weak pion production off nucleons is
dominated by the weak excitation of the $\Delta(1232)$ resonance and
its subsequent decay into $N\pi$. Thus, for $J=3/2$, isospin $I=3/2$ ($W^+ p\to\pi^+ p$),
and CM energies in the $\Delta$ region, the $L=1$ partial wave in
Eq.~(\ref{eq:watson-gral}) should be the most important. Actually, it largely
dominates the  $\pi^+p \to \pi^+ p$ reaction at
these energies for $J=3/2$. Its contribution is much larger than
the one of the $d $wave, which is also allowed.  Therefore, for the different
$r,\lambda$ values, but with fixed $M=r-\lambda$, the quantities
$\chi_{r,\lambda}(s)$, defined as (we introduce the factor $\sqrt{3\pi/8}$ for latter convenience) 
\begin{eqnarray}
\chi_{r,\lambda}(s)= \sqrt{\frac{3\pi}{8}} \sum_{\rho}(1,1/2,3/2|0,-\rho,-\rho)\,\langle
3/2,M; {0,\rho}|T(s)|0,0;{r,\lambda}\rangle,\ \ M=r-\lambda,
\label{eq:fase33}
\end{eqnarray}
should have the phase, $\delta_{P_{33}}(s)$, of the $L_{2J+1,2I+1} = P_{33}$ $\pi N$
partial wave. Expressing  the $|J M\rangle$ $\pi N$ intermediate state in terms
of helicity CM two-particle states [Eq.~(\ref{eq:cambio-base})], we
finally find
\begin{eqnarray}
\hspace*{-.75cm}\chi_{r,\lambda}(s)
e^{-i\delta_{P_{33}}}=\sqrt{\frac{3}{8}}\Bigg(\sum_{\rho}(1,1/2,3/2|0,-\rho,-\rho)\,\int
d\Omega\ {\cal D}^{(3/2)}_{M\ -\rho}(\varphi,\theta,-\varphi)\underbrace{\langle
\theta,\varphi;{0,\rho}}_{\pi^+ p}|T(s)|\underbrace{0,0;{r,\lambda}\rangle}_{W^+p}\Bigg) e^{-i\delta_{P_{33}}}
\in\reales \label{eq:watson-part}
\end{eqnarray}
for $ r=0,\pm 1,\, \lambda=\pm 1/2$ and $ M=r-\lambda$.  There appear
six, in principle, independent amplitudes. The phase of all of them
should be $\delta_{P_{33}}$.

Note that $\chi_{r,\lambda}$ in Eq.~(\ref{eq:watson-part}) is given in
terms of amplitudes between CM states with
well-defined three momenta and helicities, which could be readily
obtained in quantum field theoretical descriptions of the
$W^+p\to \pi N$ reaction, such as the HNV model presented in the
Introduction. Even for  $J=3/2$,$I=3/2$ and only $L=1$, the
HNV model does not fulfill the constraints implicit in Eq.~(\ref{eq:watson-part}).

To improve the HNV model we (partially)
unitarize it in the same fashion as in
Refs.~\cite{Carrasco:1989vq,Gil:1997bm} for pion production induced by
real and virtual photons, respectively.  We follow
the procedure suggested by M.G. Olsson in Ref.~\cite{Olsson:1974sw} and, for every given value of
the four-momentum transfer squared $q^2$,  introduce  small phases $\Psi_{V,A}(\sqrt{s}, q^2)$,
which correct the vector and axial $\Delta$ terms in the amplitude.

The matrix element
\begin{eqnarray}
\langle
\theta,\varphi;{0,\rho}|T(s)|0,0;{r,\lambda}\rangle=\epsilon_{r\mu}
T^\mu_{\lambda\rho}(\theta,\varphi)=
\epsilon_{r\mu}
T^\mu_{B\,\lambda\rho}(\theta,\varphi)+\epsilon_{r\mu}
T^\mu_{\Delta\,\lambda\rho}(\theta,\varphi),
\end{eqnarray}
can be split  into a background ($B$) and a direct Delta ($\Delta$)
contribution. Here $\epsilon_{r\mu}$ is the polarization vector of the
initial $W$ boson. We now follow Ref.~\cite{Olsson:1974sw} and implement Watson's theorem 
by modifying the above expression to
\begin{eqnarray}
\epsilon_{r\mu}
T^\mu_{B\,\lambda\rho}(\theta,\varphi)+e^{i\Psi}\epsilon_{r\mu}
T^\mu_{\Delta\,\lambda\rho}(\theta,\varphi)
\end{eqnarray}
so that
\begin{eqnarray}
\sum_{\rho}(1,1/2,3/2|0,-\rho,-\rho)\,\int d\Omega\ {\cal D}^{(3/2)}_{M\ -\rho}(\varphi,\theta,-\varphi)
\left(\epsilon_{r\mu}
T^\mu_{B\,\lambda\rho}(\theta,\varphi)+e^{i\Psi}\epsilon_{r\mu}
T^\mu_{\Delta\,\lambda\rho}(\theta,\varphi)\right),\hspace{.5cm}M=r-\lambda
\end{eqnarray}
has the right phase, $\delta_{P_{33}}(s)$. As mentioned, the phase
$\Psi$ depends on the intermediate $\Delta^{++}$ invariant mass
$\sqrt{s}$ and $q^2$.
Unfortunately, there is no single phase able to do so for all
$r,\lambda$ values. Next-to-leading contributions in the chiral
expansion, which depend  explicitly on helicities, would eventually
perturbatively restore unitarity at the price of introducing new and
uncertain low-energy constants. In addition, the resulting amplitudes would be
much more complicated and difficult to handle in Monte Carlo event
generators. The practical solution proposed here is to consider two
different Olsson phases, $\Psi_V$ and $\Psi_A$, for the vector and axial 
parts of the
transition amplitude
\begin{equation}
T=T^V-T^A\,,
\end{equation}
chosen to unitarize only the
dominant vector and axial multipoles. Note that both vector and axial
parts of these dominant $W^+p\to p \pi^+$ multipoles are required to fulfill
Watson's theorem independently. This is justified because the vector part,
which is the only one present in photo- and electropion production amplitudes, 
should satisfy Watson's theorem independently and therefore 
have the phase $\delta_{P_{33}}$. 

Using invariance under parity, the number of independent amplitudes 
can be reduced down to three vector and 
three axial ones (Appendix \ref{app:chi}) because 
\begin{eqnarray}
 \chi_{r,\lambda}= \chi_{r,\lambda}^V -
 \chi_{r,\lambda}^A 
&=& -\left( \chi_{-r,-\lambda}^V +
 \chi_{-r,-\lambda}^A \right) \label{eq:relPJ}
\end{eqnarray}
To obtain the vector and axial dominant multipoles, we rewrite the 
$|3/2,M;{L'S'}\rangle$ initial  $WN$ states in Eq.~(\ref{eq:nodepM})
in terms of the set of states commonly used in pion electroproduction~\cite{Drechsel:1992pn}.
Thus, we first couple the $WN$ orbital angular momentum to the $W$ boson 
spin $(L'\otimes 1)^{\tilde l}$ and then the 
resulting $\tilde l$ angular momentum to
the nucleon spin to get  total angular momentum states with  $J=3/2$. 
The relation between the new and old states is given in terms of Racah
coefficients (${\tilde W}$),
\begin{equation}
|J M; L'S'\rangle = \sum_{\tilde l} \sqrt{(2S'+1)(2{\tilde
    l}+1)}\ {\tilde W}(1/2,1,J,L'; S'{\tilde l}) |J M; L'{\tilde l}\rangle \label{eq:baseTD}
\end{equation}
The six independent multipoles in this basis are matrix elements of the form 
\begin{equation}
\langle {L=1,S=1/2}|T^{V,A}_{J=\frac32}(s)|{L'{\tilde l}}\rangle
\end{equation}
with $(L'=1, {\tilde l}=1,2)$ and $(L'=3, {\tilde l}=2)$ for the
vector part, and $(L'=0, {\tilde l}= 1)$ and $(L'=2, {\tilde l}=1,2)$
for the axial one. The actual relations of these multipoles with the
$\chi_{r,\lambda}^{V,A}$ amplitudes can be found in  Appendix
\ref{app:chi} [Eqs.~(\ref{eq:dominantv})-(\ref{eq:dominantfinal})]. 

\begin{figure}
\vspace*{1cm}
\resizebox{!}{14.cm}{\includegraphics{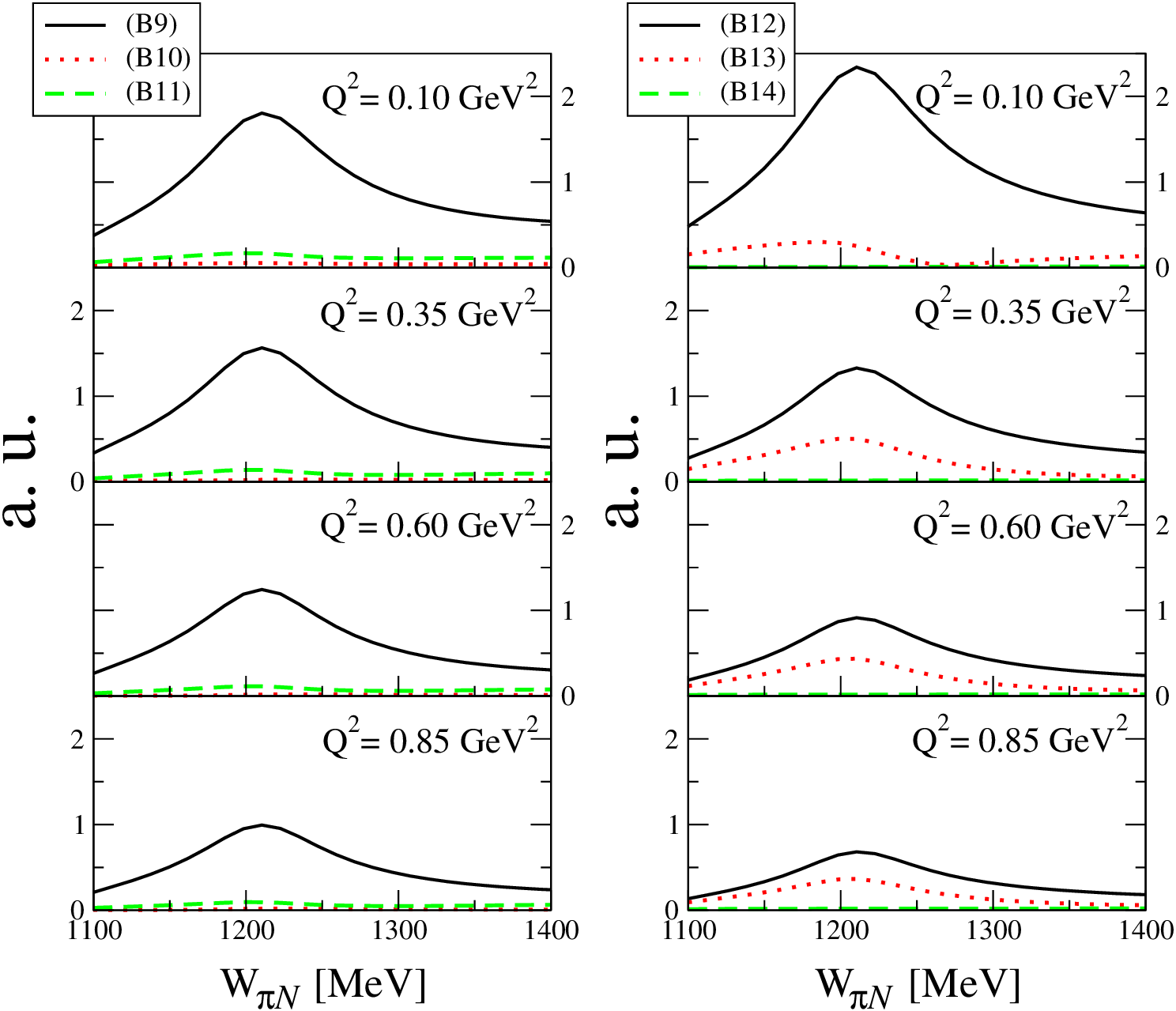}}
\caption{ Modulus of the different vector (left panels) and axial 
(right panels) multipoles defined in 
Eqs.~(\ref{eq:dominantv})-(\ref{eq:dominantfinal}) of  
Appendix~\ref{app:chi}. The scale is the same in all panels.}
  \label{fig:multipoloscomp}
\end{figure}

As discussed above, we impose Watson's theorem only
on the dominant vector and axial multipoles given, respectively, by
 Eqs.~(\ref{eq:dominantv}) 
and (\ref{eq:dominanta}). These are the magnetic $M_{1+}$ multipole in
the vector part~\cite{Drechsel:1992pn},
\footnotesize{$\langle{L=1,S=1/2}|T^{V}_{J=\frac32}(s)|{L'=1\ {\tilde l}=1}\rangle$}\normalsize, and the
$WN $ $s$-wave $\langle {L=1,S=1/2}|T^{A}_{J=\frac32}(s)|{L'=0\ {\tilde l}=1}\rangle$
multipole in the axial one. The remaining two matrix elements involve the
$WN$ pair in the relative $d$ wave ($L'=2$).  In Fig.~\ref{fig:multipoloscomp},
 we show the 
modulus of the different vector and axial multipoles defined 
in Eqs.~(\ref{eq:dominantv})-(\ref{eq:dominantfinal}) in  
Appendix~\ref{app:chi}.  The results are very similar after partial 
unitarization. From  Fig.~\ref{fig:multipoloscomp},  it is apparent that, 
while the vector multipole of Eq.~(\ref{eq:dominantv}) remains dominant 
in the whole $q^2$ range, the axial multipole of Eq.~(\ref{eq:subdominanta}) 
becomes comparable to the one of Eq.~(\ref{eq:dominanta}) as $Q^2 = -q^2$ 
increases. One might then question the approximation of imposing unitarity 
for the multipole of Eq.~(\ref{eq:dominanta}) alone in the axial sector. 
In this respect, it should be stressed that for larger $Q^2$ the contributions
 of both multipoles to the amplitudes become very similar. This is because
  the terms in which they differ (proportional to $\chi^A_{0,-1/2}$) are
   suppressed by powers of $1/\sqrt{Q^2}$ from the vector boson polarization
    for $r=0$ [Eq.~(\ref{eq:pol_r0})]. Therefore, once the dominant multipole
     of Eq.~(\ref{eq:dominanta}) fulfills Watson's theorem, it is, to a large
      degree,  also fulfilled by the subdominant one of 
      Eq.~(\ref{eq:subdominanta}).  

\begin{figure}
\vspace*{1cm}
\resizebox{!}{7.cm}{\includegraphics{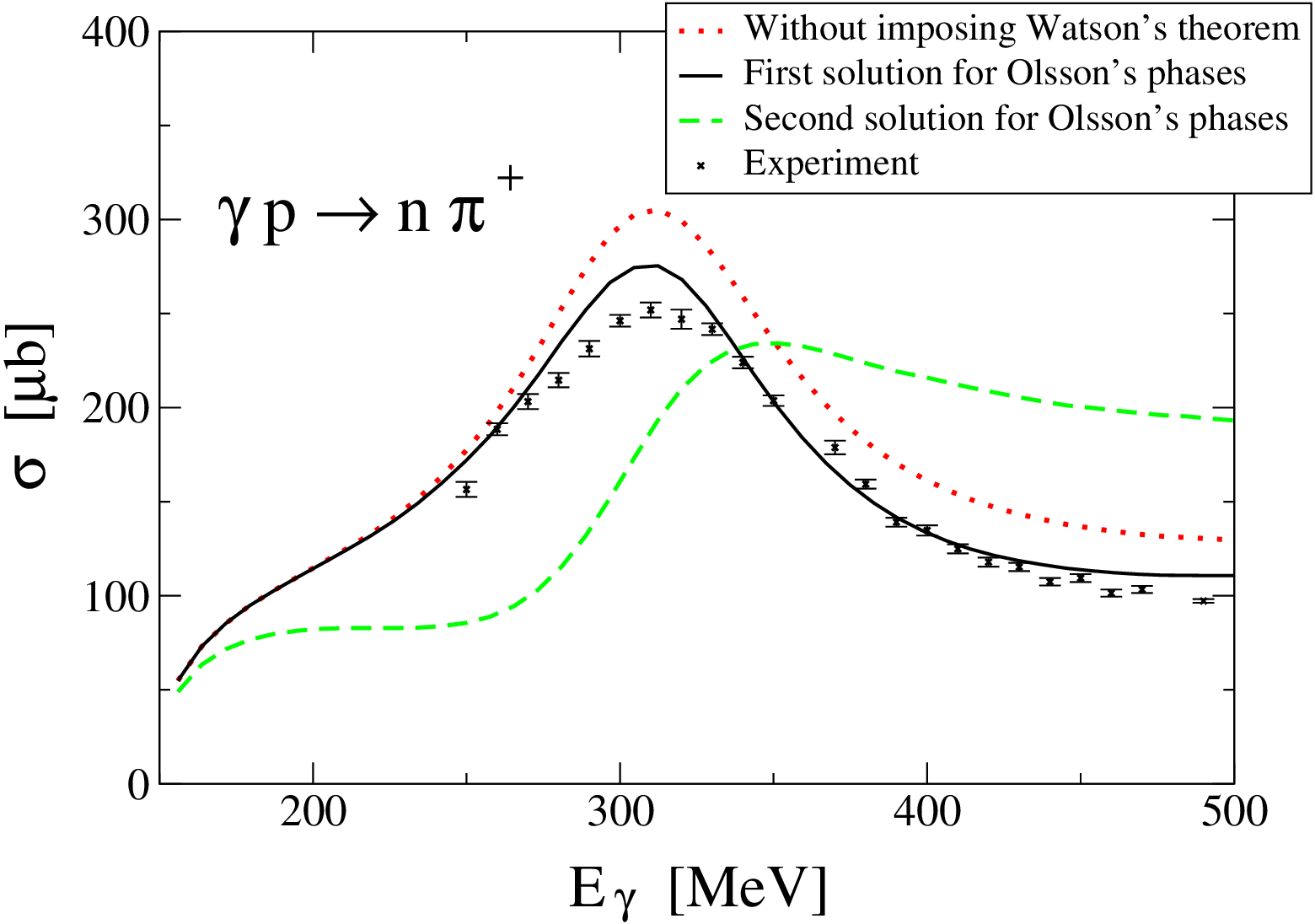}}\caption{ Results for the
$\gamma p\to n\pi^+$ reaction obtained with the vector part of our model. A better
description of the experimental data is obtained with the  smallest $\Psi_V$ 
Olsson  phase (first solution). Experimental data are taken from Ref.~\cite{Fujii:1976jg}.}
\label{fig:fotoprod}
\end{figure}

The relative $\Delta$background
phases, $\Psi_V(\sqrt{s},q^2)$ and
$\Psi_A(\sqrt{s},q^2)$,  are fixed by requiring the phase of
 each of the amplitudes $\chi^V$ and $\chi^A$, defined 
 as\footnote{Note that the symmetry
  relations of Eq.~(\ref{eq:relPJ}) guarantee that $\chi^V$
  ($\chi^A$) depends only on matrix elements of $T^V$  ($T^A$).}
\begin{eqnarray}
\chi^V &=& \frac12\left[(\chi_{1,1/2}-\chi_{-1,-1/2})+\sqrt{3}\ (\chi_{1,-1/2}-\chi_{-1,1/2})
\right]\\
\chi^A &=&
-\frac{1}{\sqrt{6}}\left[\sqrt{2}\ (\chi_{0,-1/2}+\chi_{0,1/2})+\sqrt{3}\ (\chi_{1,-1/2}+\chi_{-1,1/2})+(\chi_{1,1/2}
+\chi_{-1,-1/2})
\right]
\end{eqnarray}
to be $\delta_{P_{33}}(s)$. This is to say, we impose
\begin{equation}
{\rm Im}\left[e^{-i\delta_{P_{33}}(s)}\chi^{V,A}\right]=0. \label{eq:watsonfinal}
\end{equation}
In each case, there exist two sets of solutions, which correspond to $\chi^{V,A}$
having phases $\delta_{P_{33}}$ and $(\delta_{P_{33}}+\pi)$,
respectively (note that the $\pi N$ phase shift is
defined up to a $\pi$ factor). We take the first set of solutions, because it leads to 
the smallest $\Psi_V$ and $\Psi_A$ Olsson extra phases. The second solution for the vector current is discarded by
data on pion photoproduction off nucleons. This is shown in
Fig.~\ref{fig:fotoprod} where we apply the vector part of our model to describe
the $\gamma p\to n\pi^+$ reaction. As seen from  Fig.~\ref{fig:fotoprod}, a
much better agreement with the data is obtained when taking the solution with the
smallest $\Psi_V$  Olsson phase.
As for $\Psi_A$, the
results shown in Fig.~12 of Appendix B of Ref.~\cite{Hernandez:2007qq}
favor vector and axial $\Delta(1232)$ contributions having similar phases.

\section{Results and discussion}
\label{sec:results}
We (partially) unitarize the HNV model using   Olsson's
implementation of Watson's theorem discussed in
Sec.~\ref{sec:olss}. For this purpose, we implement the constraints implicit in
Eq.~(\ref{eq:watsonfinal}) using $\chi_{r,\lambda}$ amplitudes
calculated by means of Eq.~(\ref{eq:watson-part}). In Appendix~\ref{app:chiHNV}, 
details on the evaluation of matrix elements $\langle
\theta,\varphi;{0,\rho}|T(s)|0,0;{r,\lambda}\rangle$, which appears in
Eq.~(\ref{eq:watson-part}),  within the HNV
model are provided. For the $P_{33}$ $\pi N$ phases we have used the output of the George Washington University Partial Wave Analysis(SAID)~\cite{SAID} from which we take the WI08 single energy
values. In the analysis we neglect the influence of the small errors
(ranging from 0.1\% to 0.6\%) in
the $P_{33}$ phase shifts given in Ref.~\cite{SAID}.

\subsection{Fit A}
Following Ref.~\cite{Hernandez:2010bx}, we make a simultaneous fit to both ANL and
BNL data samples, taking into account deuterium effects, but now imposing  the unitarity of 
the two dominant multipoles $\chi^{V,A}$. This analysis gives (fit A)
\begin{eqnarray}
&&C_5^A(0)=1.12\pm0.11,\nonumber\\ 
&&M_{A\Delta}=(953.7  \pm  62.6)\,{\rm MeV}. \label{eq:bestfit}
\end{eqnarray} 
The new central value of $C_5^A(0)$ agrees within 1$\sigma$ with the
off-diagonal GTR prediction.   As
in Ref.~\cite{Hernandez:2010bx}, the
ANL~\cite{Radecky:1981fn} flux-averaged $d\sigma/dQ^2$ differential
cross section, with a $W_{\pi N}=\sqrt{s} <1.4\,$GeV cut in the final
pion-proton invariant mass, and the integrated cross sections for the
three lowest neutrino energies (0.65, 0.9 and 1.1 GeV) of the BNL
data set~\cite{Kitagaki:1986ct} have been fitted. 
A systematic error, due to flux uncertainties (20\% for ANL and 10\% for
BNL data) has been added in quadratures to the statistical one.

In Table~\ref{tab:fit}, we compare the
results for $C_5^A(0)$ and $M_{A\Delta}$ obtained in this work with
those from previous HNV fits carried out in
Refs.~\cite{Hernandez:2007qq, Hernandez:2010bx}.  With respect to the fit carried out in Ref.~\cite{Hernandez:2007qq}, the consideration of 
BNL data and flux uncertainties in  Ref.~\cite{Hernandez:2010bx} led to an increased value of $C_5^A(0)$, while strongly
reducing the statistical correlations between $C_5^A(0)$ and
$M_{A\Delta}$. The inclusion of background terms reduced $C_5^A(0)$,
while deuteron effects slightly increased it by about 5\%, consistently
with the results of Refs.~\cite{Hernandez:2007qq} and
\cite{AlvarezRuso:1998hi,Graczyk:2009qm}. The 
implementation of Watson's theorem, for the dominant vector and axial
multipoles, in new fit A,  further increases the $C_5^A(0)$ value, 
bringing it into much better agreement with the off-diagonal GTR prediction.  
%
\begin{table*}
\caption{Results from different fits to the ANL and BNL
  data. All fits  include the ANL~\cite{Radecky:1981fn}
  flux-averaged $d\sigma/dQ^2$ differential cross section, with a
  $W_{\pi N}=\sqrt{s} <1.4\,$GeV cut, and the integrated cross sections for
  the three lowest neutrino energies (0.65, 0.9 and 1.1 GeV) of the
  BNL data set~\cite{Kitagaki:1986ct}.   Fits I$^*$, II$^*$
  and IV are taken from Ref.~\cite{Hernandez:2010bx}. In all cases
  Adler's constraints ($C_3^A=0,\;C_4^A=-C_5^A/4$)
  \cite{Adler:1968tw,Bijtebier:1970ku} are imposed.  Deuteron
  effects~\cite{Hernandez:2010bx}   are included  in  fit
  IV and in those carried out in this work.  The nonresonant chiral
  background contributions are included in all cases, with the
  exception of fit I$^*$. For $C_5^A(q^2)$ a dipole form, $C_5^A(q^2)
  = C_5^A(0)/(1-q^2/M^2_{A\Delta})^2$, has been used in all fits
  except in the one carried out in Ref.~\cite{Hernandez:2007qq}, where an
  extra factor $1/(1-q^2/3M_{A\Delta})$ was included [see Eq.~(48) of
  that reference].  Finally, $r$ is the Gaussian correlation
  coefficient between $C_5^A(0)$ and $M_{A\Delta}$.  For reference,
  the prediction of the GTR is $C_5^A(0) = 1.15-1.2$.  }
\vspace{.3cm}
\begin{tabular}{c|cc|cc|c}
\hline\hline
& $C_5^A(0)$ & $M_{A\Delta}$/GeV & Data & $r$
  &$ \chi^2$/dof\\\hline  
Ref.~\cite{Hernandez:2007qq} &  $0.867\pm 0.075$ & $0.985\pm 0.082$&
 ANL & $-0.85$ & 0.40 \\ \hline
Ref.~\cite{Hernandez:2010bx}: Fit I$^*$ (only $\Delta$ pole) &  $1.08\pm 0.10$ & $0.92\pm 0.06$&
ANL \& BNL &  $-0.06$ & 0.36\\
Fit II$^*$ &  $0.95\pm 0.11$ & $0.92\pm 0.08$&
ANL \& BNL &  $-0.08$ & 0.49\\ 
Fit IV (with deuteron effects)&  $1.00\pm 0.11$ & $0.93\pm 0.07$&
ANL \& BNL &  $-0.08$ & 0.42 \\ \hline
This work (unitarized + deuteron effects) fit A& $1.12\pm 0.11$ & $0.954\pm 0.063$&
 ANL \& BNL & $-0.08$ & 0.46 \\
\hline\hline
\end{tabular}
\label{tab:fit}
\end{table*}

%

%
\begin{figure}
\vspace*{1cm}
\resizebox{!}{4.25cm}{\includegraphics{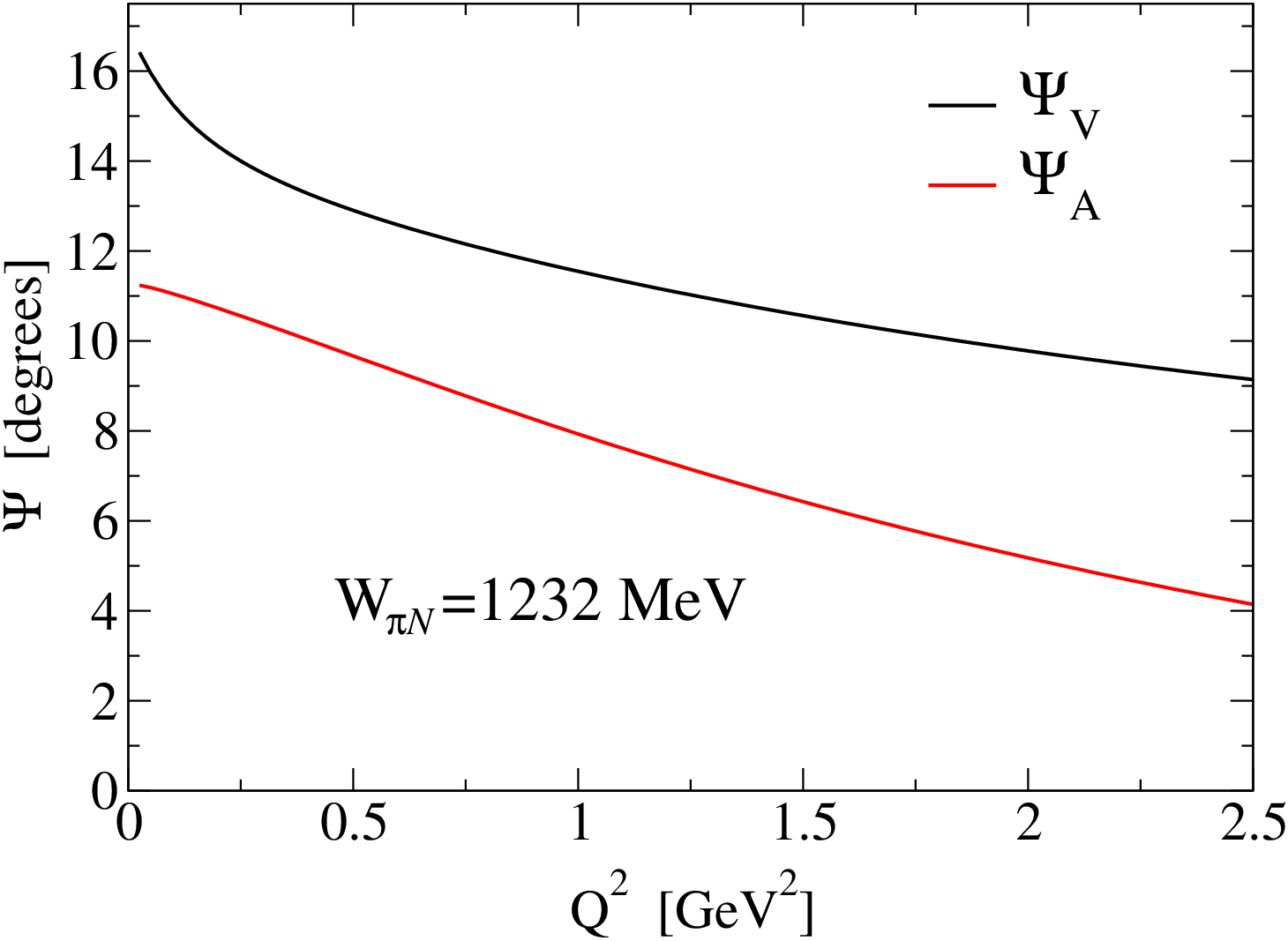}}\hspace*{.15cm}
\resizebox{!}{4.25cm}{\includegraphics{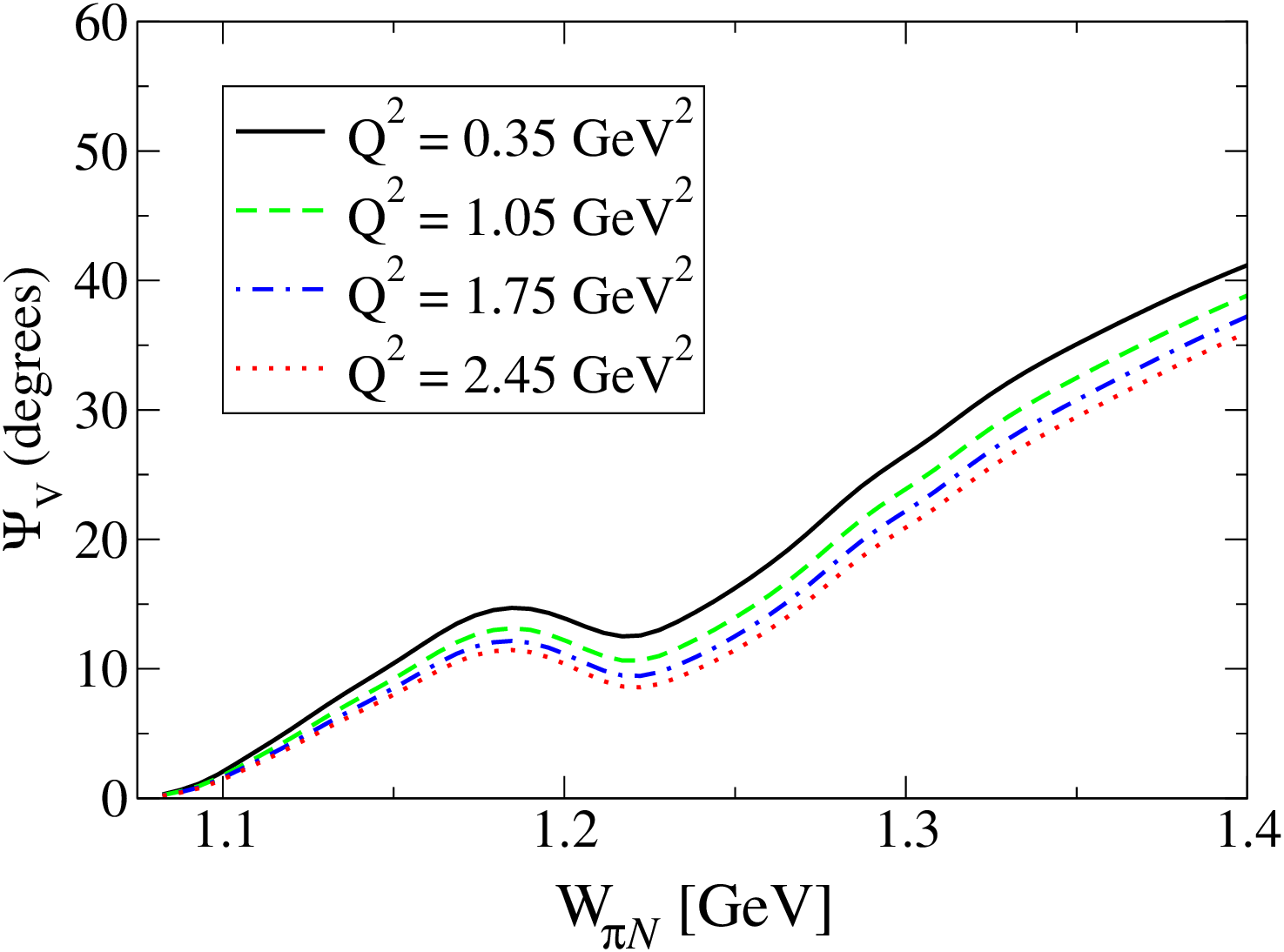}}\hspace*{.15cm}
\resizebox{!}{4.25cm}{\includegraphics{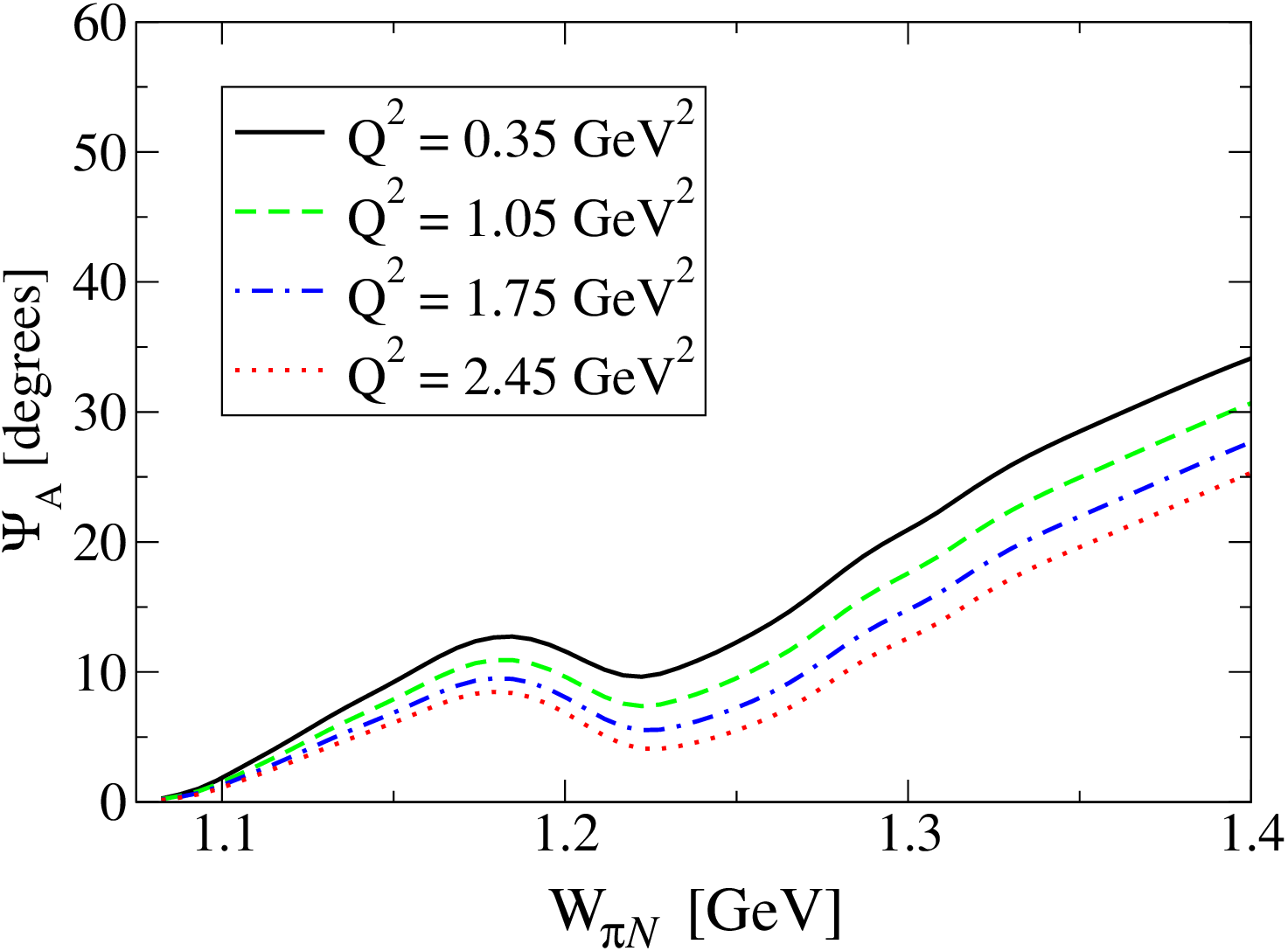}}\caption{ Olsson phases
from fit A. Left panel: 
$\Psi_V$ and $ \Psi_A$ at
the $\Delta$ peak as a function of $Q^2=-q^2$. 
Middle and right
panels: $\Psi_V$ and $ \Psi_A$ as a function of 
 the $\Delta$ invariant mass $W_{\pi N}$ for different $Q^2$ values, 
 respectively.}
  \label{fig:fases}
\end{figure}
The resulting Olsson phases from fit A
 are depicted in Fig.~\ref{fig:fases}. In the left panel of Fig.~\ref{fig:fases} we show the phases
$\Psi_V, \Psi_A$ obtained at the $\Delta$ peak as a function of
$Q^2$. 
In the middle and right panels of Fig.~\ref{fig:fases}
we give, for different $Q^2$ values, the $\Psi_V, \Psi_A$ dependence
on the $\Delta$ invariant mass $W_{\pi N}$.   The vector 
phase $\Psi_V$ agrees reasonably well with the one 
determined for electron scattering in Ref.~\cite{Gil:1997bm}.

\begin{figure}[t]
\vspace{.5cm}\resizebox{!}{5cm}{\includegraphics{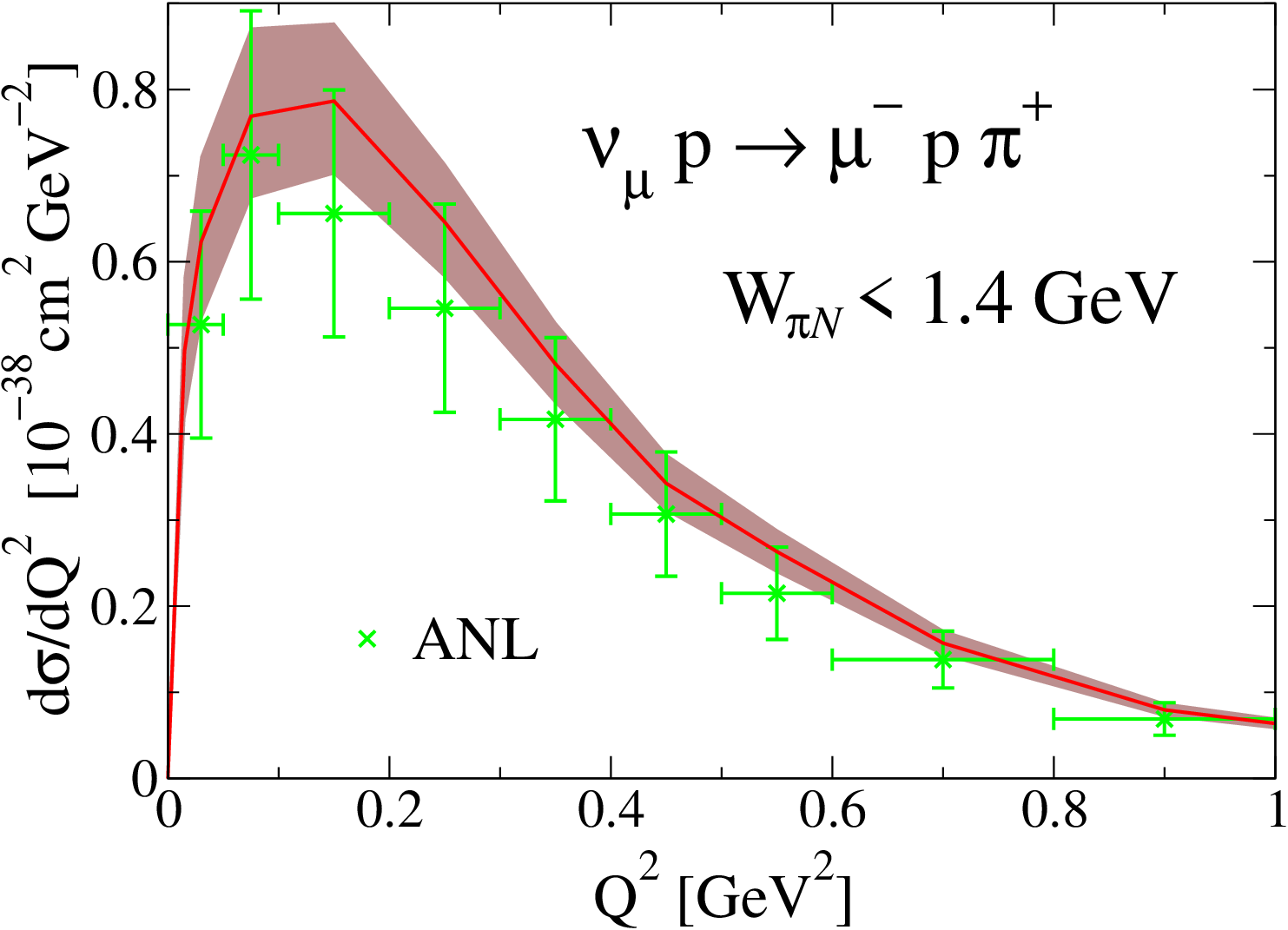}}\hspace*{.15cm}
\resizebox{!}{5cm}{\includegraphics{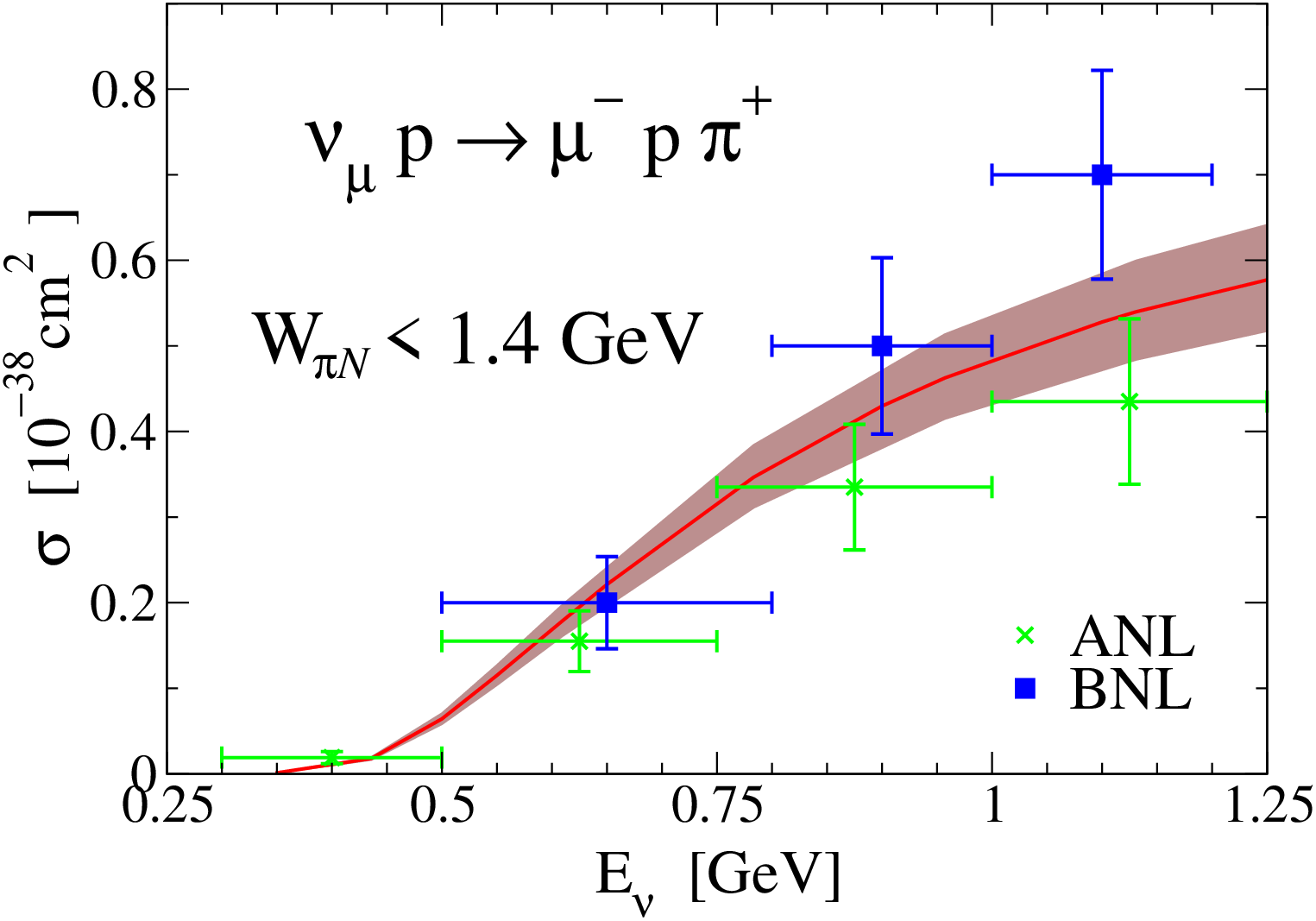}}\\
\caption{ Results from fit A  for the
  differential $d\sigma/dQ^2$ (left) and total (right) cross section for 
  the $\nu_\mu p\to \mu^- p\pi^+$ reaction compared to the 
  ANL~\cite{Radecky:1981fn}~and~BNL~\cite{Kitagaki:1986ct} data.   
  The  
  theoretical bands correspond to the variation of the results
 when $C_5^A(0)$ changes within the error interval determined from the fit.
Experimental data include a systematic error, due to flux uncertainties (20\% for ANL
  and 10\% for BNL data), which has been added in quadratures to the
  statistical ones. Theoretical results and ANL data include a cut in
  the final pion-proton invariant mass given by $W_{\pi N}< 1.4\,$GeV. Deuteron effects have been taken into account assuming that the neutron in the deuteron acts
  as a spectator (details in Ref.~\cite{Hernandez:2010bx}).
}
\label{fig:fit}
\end{figure}

The results of the (partially) unitarized model derived in this work (fit A) 
are confronted to the fitted data in Fig.~\ref{fig:fit}. The same good 
agreement to the data as in Ref.~\cite{Hernandez:2010bx}, where partial unitarity was not imposed, is now obtained with a higher $C_5^A(0)$ consistent with the GTR.
The increase in the $C_5^A(0)$ value with respect to that calculation 
is compensated by the change in the interference  between the dominant
$\Delta$ term and the background terms once Watson's theorem is imposed on the
dominant multipoles.

\begin{figure}
\resizebox{!}{5cm}{\includegraphics{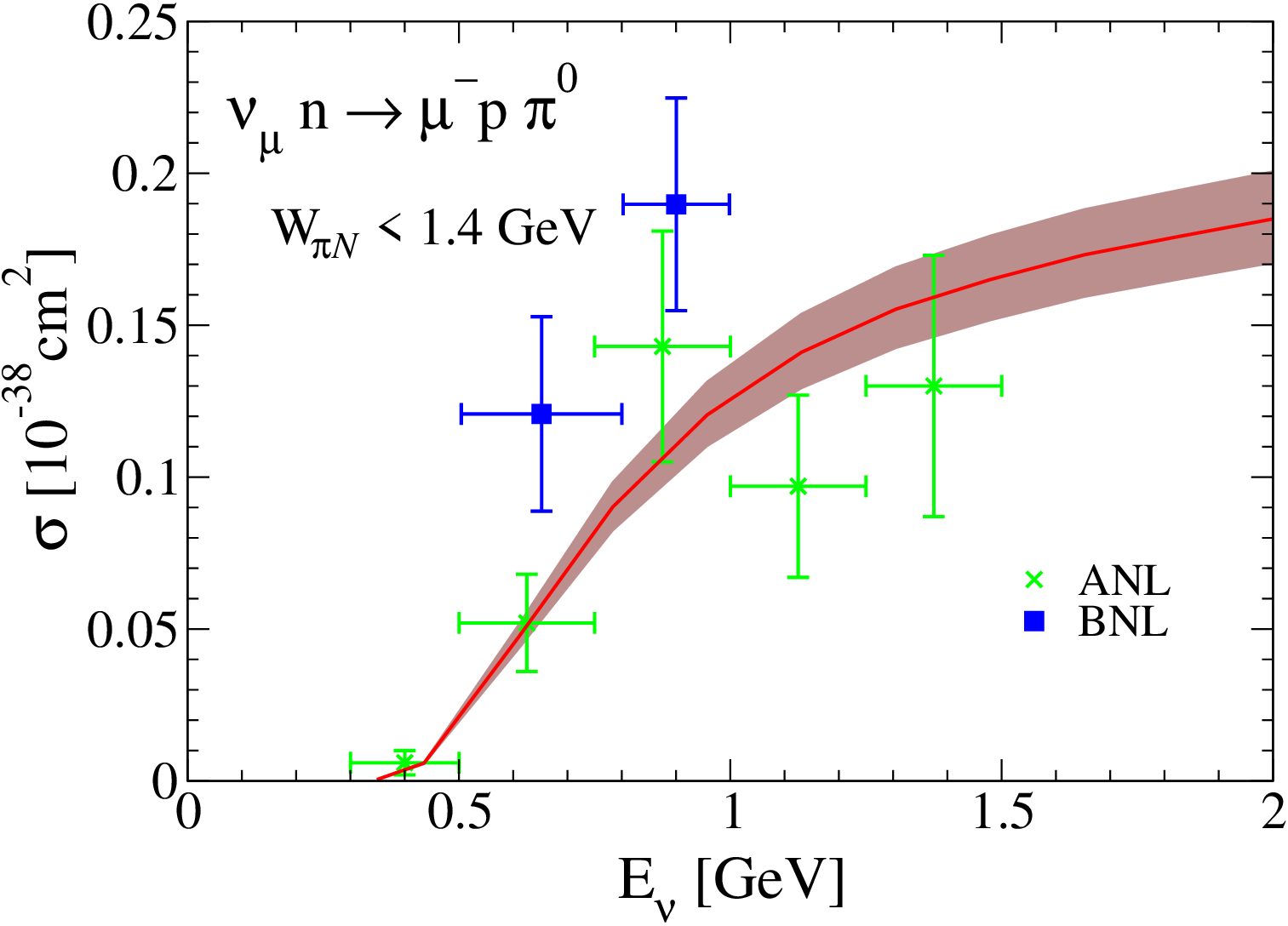}}\hspace*{.15cm}
\resizebox{!}{5cm}{\includegraphics{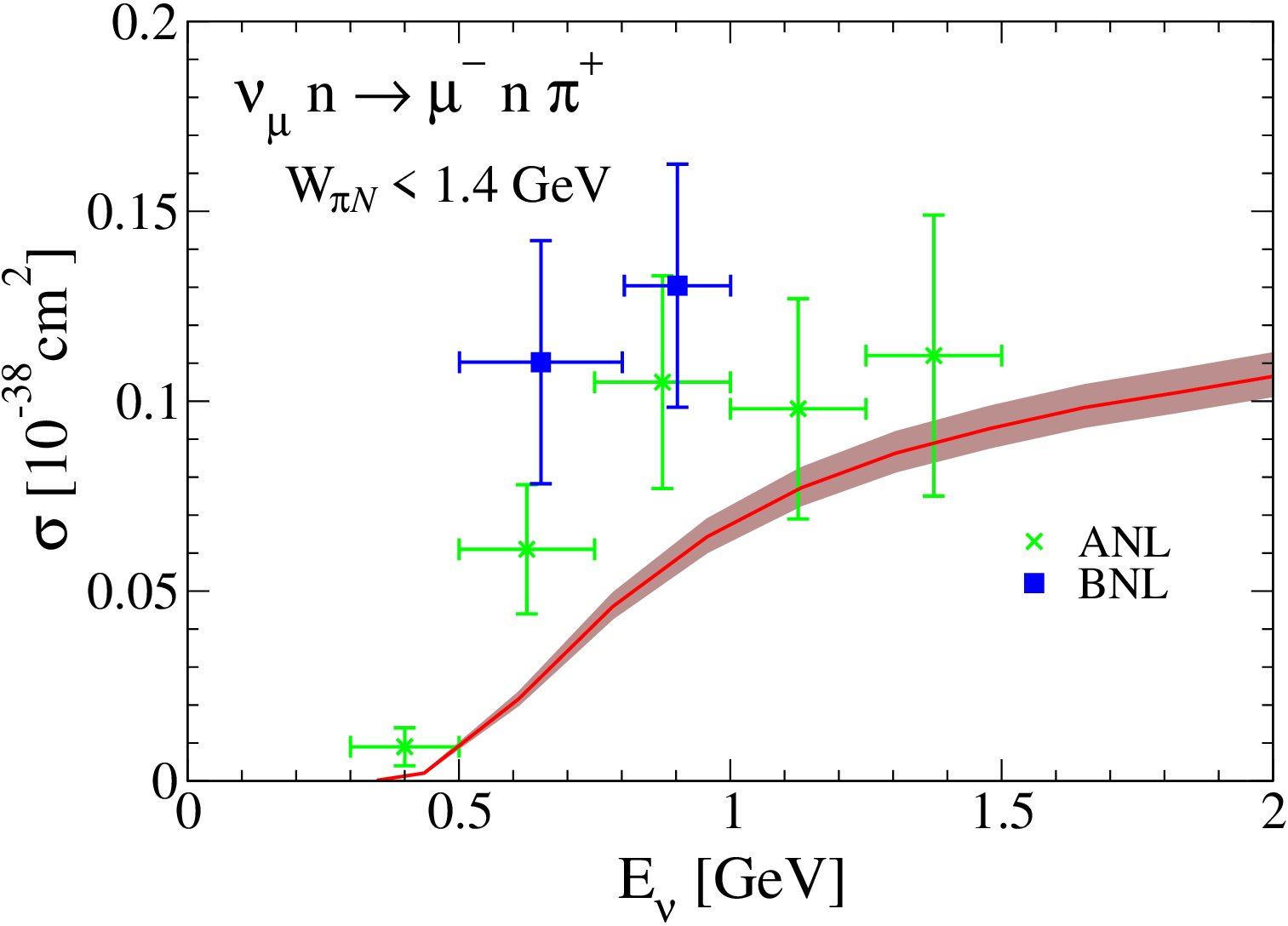}}
\caption{ Results from fit A [Eq.~(\ref{eq:bestfit})] for the
  $\nu_\mu n\to \mu^- p\pi^0$ (left panel) and
  $\nu_\mu n\to \mu^- n\pi^+$ (right panel) total cross sections as compared to
  ANL~\cite{Radecky:1981fn} and BNL~\cite{Kitagaki:1986ct} data.
  Theoretical bands and experimental errors have the same meaning as in
  Fig.~\ref{fig:fit}. Theoretical results and ANL data include a cut
  in the final active nucleon-pion invariant mass given by $W_{\pi N}
  <1.4\,$GeV. Deuteron effects have been taken into account as
  explained in Ref.~\cite{Hernandez:2010bx}, assuming that the proton in the deuteron acts as a spectator.\vspace{.35cm} }
\label{fig:n}
\end{figure}

In Fig.~\ref{fig:n} we show the predictions of the partially
unitarized (Fit A) HNV model for the $\nu_\mu n\to \mu^- p\pi^0$ and
$\nu_\mu n\to \mu^- n\pi^+$ channels. They are compared to the ANL
and BNL data, assuming that the proton in the deuteron acts as a
spectator. The problem with the $\nu_\mu n\to \mu^- n\pi^+$ channel,
where data are underestimated in most theoretical models, still persists
after partial unitarization. This significant discrepancy deserves
additional work, even more so because there exist only two independent
amplitudes, and thus the $p\pi^0$ and $p\pi^+$ channels fully
determine the $n\pi^+$ amplitude~\cite{Hernandez:2007qq}. We would
like to point out that the crossed $\Delta$ mechanism has a large
contribution in the $n\pi^+$ channel. Indeed, besides the
$\Delta$ propagator, the numerical factors of the (direct and crossed)
$\Delta$ mechanisms are $(\sqrt{3}~ \&~ 1/\sqrt{3})$, $(2/\sqrt{3} ~
\&~ -2/\sqrt{3})$ and $(1/\sqrt{3} ~ \&~ \sqrt{3})$ for the $p\pi^+$,
$p\pi^0$ and $n\pi^+$ channels,
respectively~\cite{Hernandez:2007qq}. The spin structure of the
$\Delta$ propagator used in Ref.~\cite{Hernandez:2007qq} suffers from some
off-shell ambiguities/inconsistencies, which are clearly enhanced in
the evaluation of the crossed term, where the resonance is far
from its mass shell. This might have consequences, which would affect
much more the $n\pi^+$ channel than the other two charge
configurations. Research along these lines is underway.

Effects of the final state interactions (FSIs) on
cross sections for the single pion production off the deuteron should
also be  considered and might help to explain the puzzling
$n\pi^+$ channel. Such effects have been recently examined in the
work of Ref. \cite{Wu:2014rga}. There, it is found that the
orthogonality between the deuteron and final $pn$ scattering wave
functions significantly reduces the cross sections. Thus the ANL and
BNL data on the deuterium target might need a more careful analysis with
the FSIs taken into account. It is also relevant to incorporate
 the kinematical cuts implemented in the 
experiments to properly separate the three reaction channels.

\begin{figure}[]
\resizebox{!}{5cm}{\includegraphics{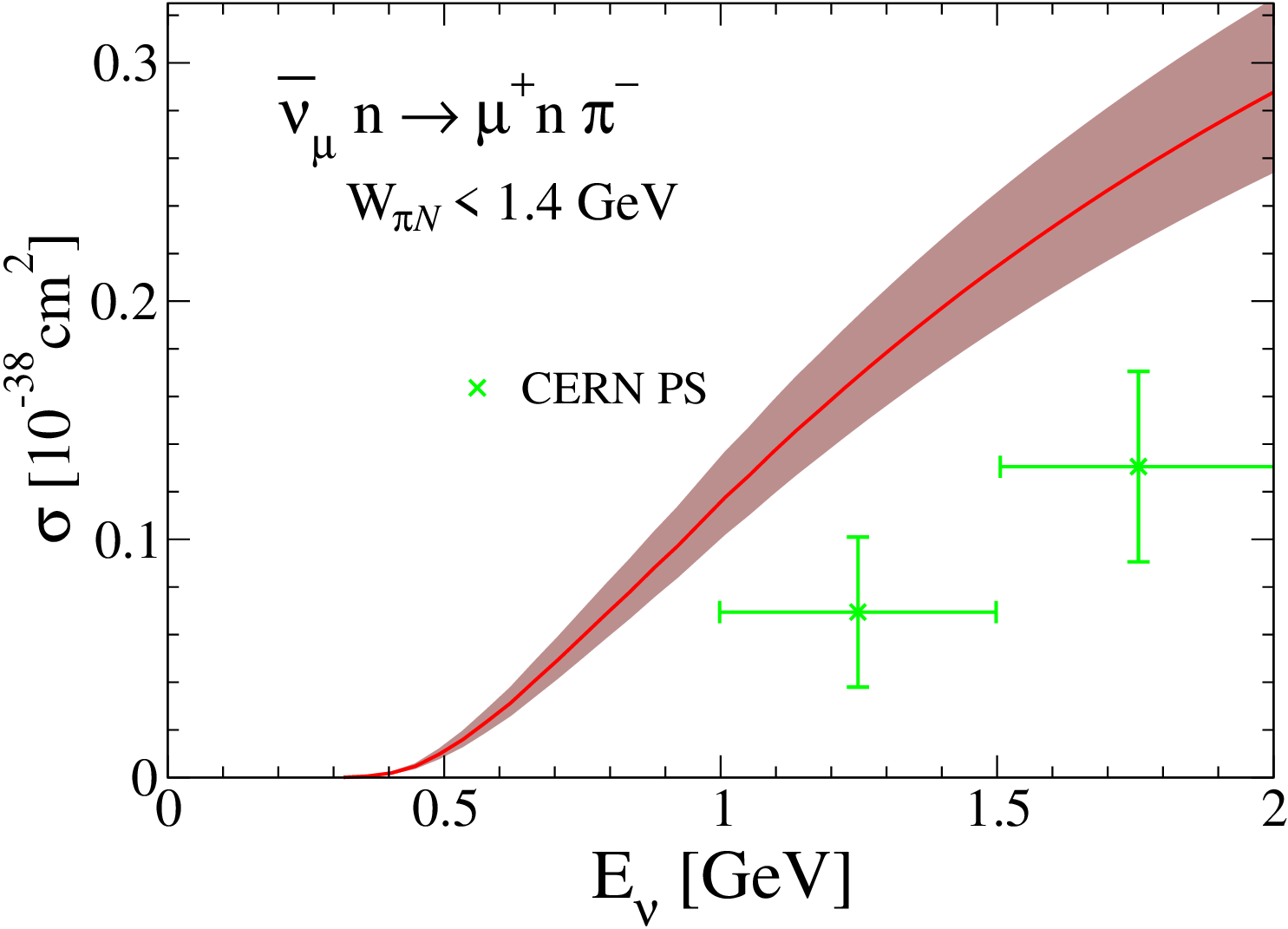}}\hspace*{.25cm}
\resizebox{!}{5cm}{\includegraphics{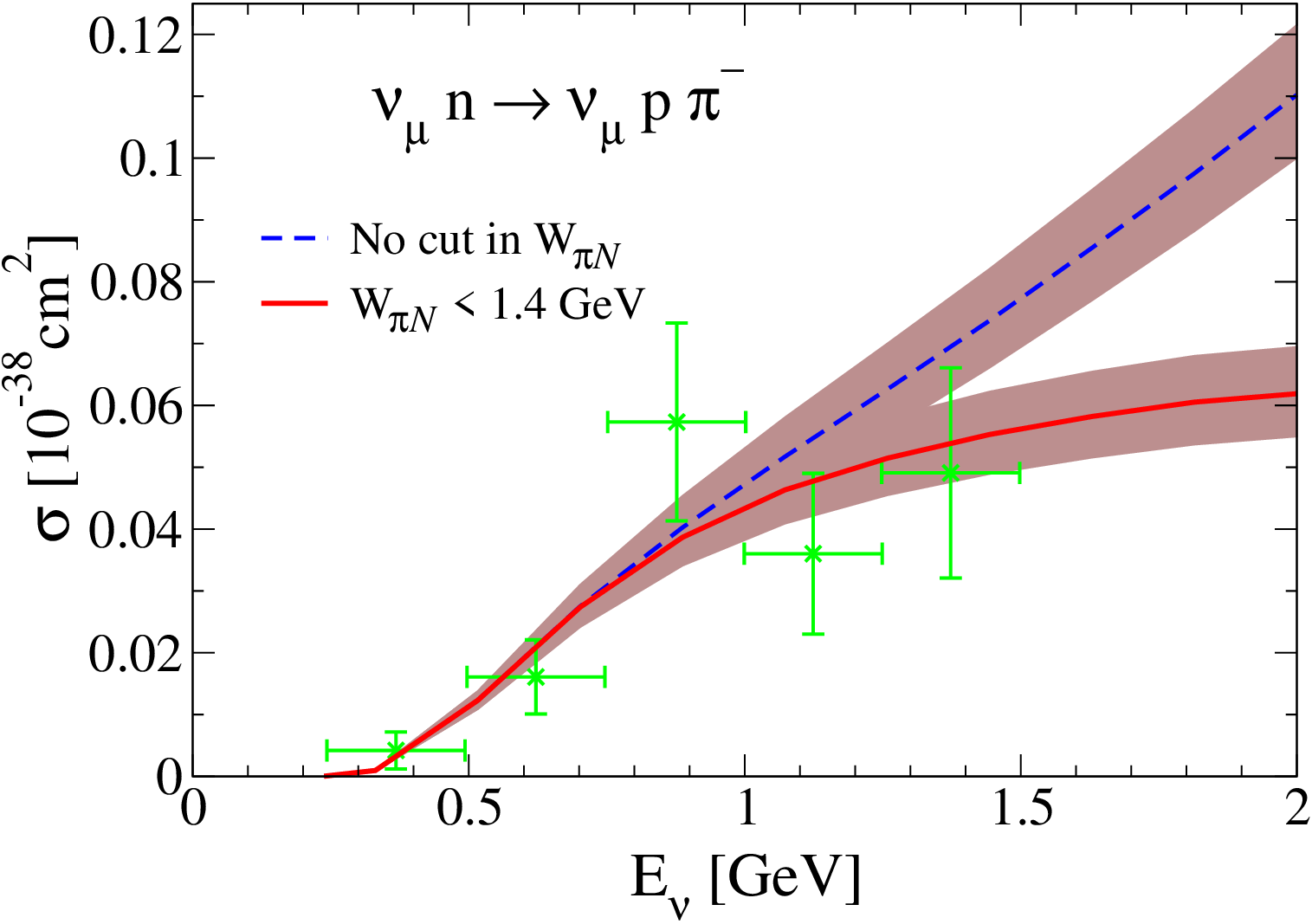}}
\caption{Results from fit A [Eq.~(\ref{eq:bestfit})] for the $\bar\nu_\mu n\to\mu^+n\pi^-$ (left panel)
and $\nu_\mu n\to \nu_\mu p\pi^-$ (right panel) total
cross sections. Data from CERN PS were taken in a freon-propane
($\rm CF_3Br-C_3H_8$) target~\cite{Bolognese:1979gf}. Experimental data for the
$\nu_\mu n\to \nu_\mu p\pi^-$ reaction are from Ref.~\cite{Derrick:1980nr}.
Since this  latter cross section was measured at ANL, we have assumed a 20\% systematic error, due to flux
uncertainties, that has been added in quadratures to the statistical error.
Besides, for the $\nu_\mu n\to \nu_\mu p\pi^-$ case, we have taken into account deuteron effects, as explained in Ref.~\cite{Hernandez:2010bx}, assuming the proton in the deuteron as a spectator. Theoretical 
bands have the same meaning as in Fig.~\ref{fig:fit}. Theoretical results, except where indicated,  include a cut in the final pion-nucleon
invariant mass given by $W_{\pi N} < 1.4\,$GeV. 
}
\label{fig:extra}
\end{figure}

Finally, in Fig.~\ref{fig:extra} we give the fit A results for the
$\bar\nu_\mu n\to\mu^+n\pi^-$ and $\nu_\mu n\to \nu_\mu p\pi^-$
channels. In the first case we compare with the data from
Ref.~\cite{Bolognese:1979gf} that were obtained at the CERN proton synchroton (PS) using a
freon-propane ($\rm CF_3Br-C_3H_8$) target. There is a large
discrepancy in this case between the theoretical calculation and the
experimental data.  As shown in Ref.~\cite{Athar:2007wd}, this can be
explained by nuclear medium and pion absorption effects, which 
were not properly taken into account in the analysis of
Ref.~\cite{Bolognese:1979gf}. For the second reaction, we find a nice
agreement with the experimental data from Ref.~\cite{Derrick:1980nr}.

\subsection{Fit B}

The ANL and BNL bubble chamber pion production measurements have been 
recently revisited~\cite{Wilkinson:2014yfa}. Both experiments have been
reanalyzed to produce the  ratio between the 
$\sigma(\nu_\mu p\to\mu^- p\pi^+)$ and the charged current quasielastic 
(CCQE) cross sections measured in deuterium,
cancelling in this way  the flux uncertainties present in the data. A good
agreement between the two experiments for these ratios was found, providing in 
this way an
explanation to the longstanding  tension between the two data sets. 
By multiplying the cross section ratio by the theoretical CCQE cross
section on the 
deuteron\footnote{They use the 
prediction from GENIE 2.9~\cite{Andreopoulos:2009rq}.}, 
which is well under control, flux normalization independent
pion production cross sections were extracted. We have taken advantage of these developments
and performed a new fit considering some of the new data points.  

\begin{figure}
\vspace{.35cm}
\resizebox{!}{5cm}{\includegraphics{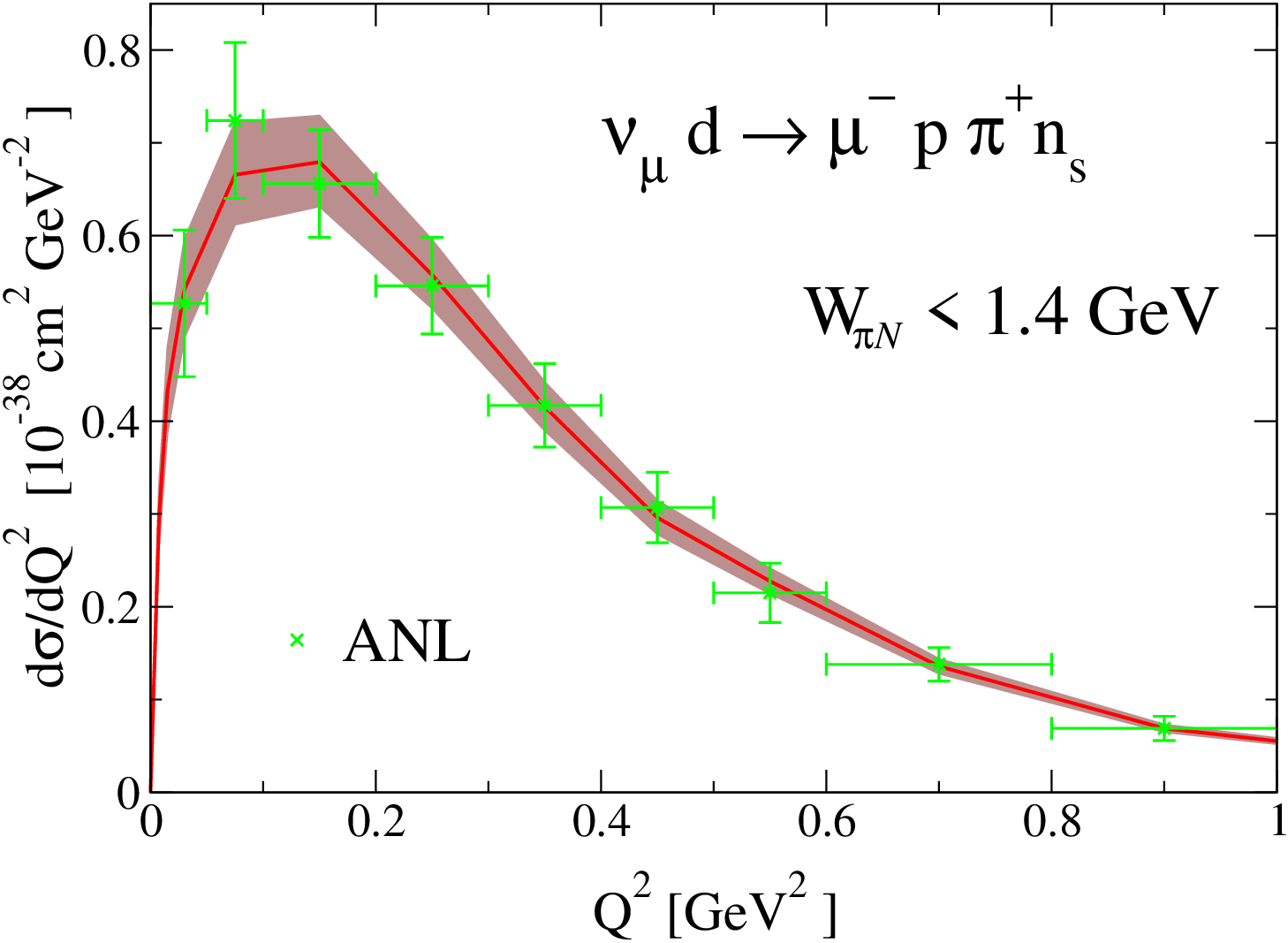}}\hspace*{.15cm}
\resizebox{!}{5cm}{\includegraphics{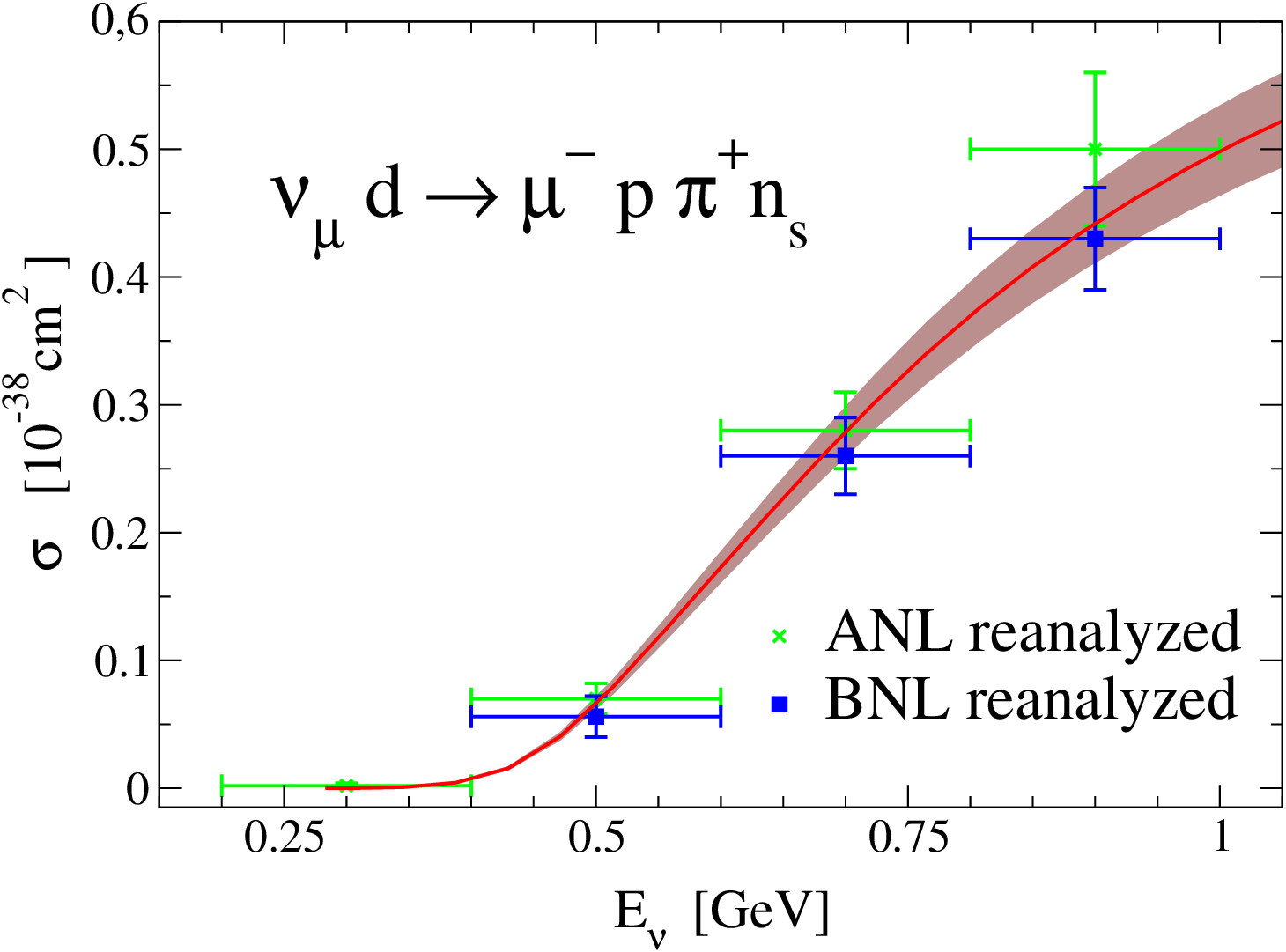}}
\caption{ Results from fit B [Eq.~(\ref{eq:bestfit2})] for the
 shape of the differential $d\sigma/dQ^2$ (left) and for total  cross 
 sections (right) for the $\nu_\mu p\to \mu^- p\pi^+$ reaction compared , respectively, to the 
  ANL~\cite{Radecky:1981fn} data and the ANL and BNL reanalyzed total 
  cross sections of Ref.~\cite{Wilkinson:2014yfa}. Theoretical bands are as in Fig.~\ref{fig:fit}. In the left panel, both experimental data and theoretical results include a cut in the final pion-neutron 
  invariant mass given by $W_{\pi N} <1.4\,$GeV.
  Besides, theoretical results in the
  left panel have been divided by $\beta=1.19$, accounting for flux uncertainties [see Eq.(\ref{eq:newchi})]. Deuteron effects have been taken into account as
  explained in Ref.~\cite{Hernandez:2010bx}.}
\label{fig:fitB}
\end{figure}

We have minimized 
\begin{equation}
\chi^2= \sum_{i\in {\rm ANL}}\left ( \frac{\beta d\sigma/dQ_i^2|_{\rm
    exp}-d\sigma/dQ_i^2|_{\rm th}}{\beta \Delta(d\sigma/dQ_i^2|_{\rm
    exp}) }\right)^2+  \sum_{i\in {\rm ANL}}\left ( \frac{\sigma_i|_{\rm
    exp}-\sigma_i|_{\rm th}}{ \Delta(\sigma_i|_{\rm
    exp}) }\right)^2+  \sum_{i\in {\rm BNL}}\left ( \frac{\sigma_i|_{\rm
    exp}-\sigma_i|_{\rm th}}{ \Delta(\sigma_i|_{\rm
    exp}) }\right)^2\label{eq:newchi} \,.
\end{equation}
The ANL and BNL integrated cross sections included in the above
$\chi^2$, taken from Ref.~\cite{Wilkinson:2014yfa}, are
  collected in Table~\ref{tab:anl-bnl-data}. 
\begin{table*}
\caption{ ANL and BNL integrated cross sections (in units of $10^{-38}$ cm$^2$) taken from the
  reanalysis of Ref.~\cite{Wilkinson:2014yfa} and included in the
  $\chi^2$ of Eq.~(\ref{eq:newchi}) (fit B).  }
\vspace{.3cm}
\begin{tabular}{c|cc|c}
\hline\hline
 $E_\nu$ (GeV) & $\sigma|_{\rm
    exp}$ & $\Delta(\sigma|_{\rm
    exp})$ & Exp. \\\hline  
0.3 & 0.0020 & 0.0020 & ANL \\
0.5 & 0.070 & 0.012 & ANL \\
0.7 & 0.28 & 0.03 & ANL\\
0.9 & 0.50 & 0.06 & ANL\\
\hline
0.5 & 0.056 & 0.016 & BNL \\
0.7 & 0.26 & 0.03 & BNL \\
0.9 & 0.43 & 0.04 & BNL \\
\hline\hline
\end{tabular}
\label{tab:anl-bnl-data}
\end{table*}
%
%
%
Since no cut in the outgoing
pion-nucleon invariant mass was considered in the new analysis of Ref.~\cite{Wilkinson:2014yfa}, and in order to avoid heavier resonances from playing a significant role, we have
only included data points corresponding to laboratory neutrino
energies $E_\nu \le 1.1$ GeV. To constrain the $q^2$ dependence, we
have also fitted  the shape of the original ANL flux-folded $d\sigma/dQ^2$
distribution, not affected by the new analysis of
Ref.~\cite{Wilkinson:2014yfa}, where  a $W_{\pi N}=\sqrt{s} <1.4\,$GeV cut in the final
pion-proton invariant mass was implemented. The new best fit parameter $\beta$ in
the first term of Eq.~(\ref{eq:newchi}) is an arbitrary
scale that allows us to  consider only the shape of this 
distribution. In turn, we do not now include any systematic error on the ANL
$d\sigma/dQ^2$  differential cross section. As in fit A, we consider deuterium
 effects and Adler's constraints ($C_3^A=0,\;C_4^A=-C_5^A/4$)
  on the axial form factors and for $C_5^A(q^2)$ use the dipole
  functional form   shown in the 
 caption of
 Table~\ref{tab:fit}. Besides, Olsson's approximate
 implementation of Watson's theorem is also taken into account.  
The best fit parameters in this case (fit B) are
\begin{eqnarray}
C_5^A(0)&=&1.14\pm0.07,\nonumber\\ 
M_{A\Delta}&=&(959.4  \pm  66.9)\,{\rm MeV}, \label{eq:bestfit2}
\end{eqnarray}
with $\beta=1.19\pm 0.08$ and $\chi^2/dof=0.3$.  
The values for $C_5^A(0)$ and $M_A$ from fit B are very close to the 
ones obtained in fit A. 
Without including the Olsson phases,  fit B gives
a  smaller $C_5^A(0)=1.05\pm0.07$  value, in worse
agreement with the GTR prediction. This is the same effect seen when 
comparing 
fit A with fit IV in Ref.~\cite{Hernandez:2010bx}.
 The value $\beta=1.19\pm 0.08$ suggests  that ANL results in 
 Ref.~\cite{Radecky:1981fn} could have underestimated the pion production cross 
 sections by some $20\%$ due to neutrino flux uncertainties. A comparison 
 of the theoretical results from fit B and the  fitted data is now shown in
 Fig.~\ref{fig:fitB}. Similar results to those from fit A are obtained for the 
 Olsson phases and the cross sections for the other channels. 

\section{Final remarks}

Pion production on deuteron target induced by neutrinos and antineutrinos 
has been studied using the HNV model~\cite{Hernandez:2007qq}, which takes into 
account nonresonant amplitudes,
 required by chiral symmetry, as well as resonant ones with $\Delta(1232)$ 
 and $N^*(1520)$ intermediate states. Phenomenological form factors allow 
 us to apply the model to finite 4-momentum transfers $q^2$ probed in neutrino
  experiments. The model has now been improved by imposing Watson's theorem 
  to the dominant vector and axial multipoles. In this way, unitarity has 
  been partially restored.

With this theoretical tool we have undertaken a new determination of the 
leading axial $N\Delta(1232)$ transition form factor from ANL and BNL data. 
We have fitted not only the original data (fit A) but also those obtained 
in a recent reanalysis~\cite{Wilkinson:2014yfa} that has removed the 
tension between the two data sets by considering flux independent ratios 
(fit B). Both fits A and B show that the partial unitarization increases
the value of the leading axial coupling $C_5^A(0)$ with respect to fits where 
no unitarization was applied. Thanks to the new analysis of 
Ref.~\cite{Wilkinson:2014yfa}, the error in $C_5^A(0)$ has been reduced 
from 10\% (fit A) to 6\% (fit B).
The agreement with the data is equally satisfactory as in previous fits 
performed without unitarization, but the new  $C_5^A(0)$ values are in better 
agreement with the prediction from the off-diagonal GTR. One should also 
mention that the description of pion photoproduction at the $\Delta(1232)$ 
peak is also improved without refitting the electromagnetic couplings 
(Fig.~\ref{fig:fotoprod}). It is the new interference pattern between 
the $\Delta$-pole amplitude  and background contributions that compensates 
for the increase in the $C_5^A(0)$ value. 
Actually, the results are compatible with the ones obtained in a
simpler model where only the dominant $\Delta$ mechanism was included
and where $C_5^A(0)\approx 1.15-1.2$, as given by the off-diagonal
GTR.  However, a more complete model containing not only the $\Delta$ 
mechanism but also background terms is definitely more robust. In fact, 
 as shown in
Ref.~\cite{Hernandez:2007qq}, there are parity violating observables
that are nonzero only in the presence of background terms. 

Full unitarity is also to be preferred. The advantage of the simpler 
scheme adopted here resides mostly in its simplicity. This would allow
 for an easier implementation in
event generators used in the analysis of neutrino experiments while, at 
the same time, providing an accurate description of the pion production 
data for $W_{\pi N}< 1.4\,$GeV. The framework is also general enough to 
correct for deviations from  Watson's theorem in more elaborated weak 
pion production models. The accuracy can be also increased by fixing the 
phases in other subdominant multipoles.

\begin{acknowledgments}
We thank Callum Wilkinson for making the results of 
Ref.~\cite{Wilkinson:2014yfa} available to us. This research
 has been supported by the Spanish Ministerio de Econom\'\i a y
 Competitividad (MINECO) and the European fund for regional development (FEDER)  under Contracts
 FIS2011-28853-C02-01, FIS2011-28853-C02-02, FPA2013-47443-C2-2-P, FIS2014-51948-C2-1-P,
 FIS2014-51948-C2-2-P,  FIS2014-57026-REDT and SEV-2014-0398 , by
 Generalitat Valenciana under Contract PROMETEOII/2014/0068 and by the
 European Union HadronPhysics3 project, grant agreement no. 283286.
\end{acknowledgments}

\appendix
\section{CM two-particle helicity states}
\label{sec:states}

We follow the notation in Ref.~\cite{martin1970elementary}, up to some
trivial factors in the normalization of the states. Particle states
are defined by the Poincar\'e symmetry group Casimir operators.  Thus,
the states $|m,j,\vec{p},\lambda\rangle$ are characterized by the mass
($m$), spin ($j$), 3-momentum ($\vec{p}$\,) and
helicity\footnote{Spin component along the direction of motion.} ($\lambda$) of the particle. They are constructed as
\begin{equation}
|m,j;\vec p,\lambda\rangle=R(\varphi,\theta,-\varphi) Z_{|\vec{p}\,|}|m,j;\vec 0,\lambda\rangle \,,
\end{equation}
with $Z_{|\vec{p}\,|}$ being a boost in the positive $z$ direction and
$R(\varphi,\theta,-\varphi)$ a rotation that takes that axis into the
direction of $\vec p$ ($\theta,\varphi$ are the polar and azimuthal 
angles of $\vec{p}$, $0\le \theta \le \pi$, $0\le \varphi < 2\pi$). 
The state $|\vec 0,\lambda\rangle$ has $\vec p=\vec 0$ and spin projection 
along the Z axis $\lambda$.
After the transformations, $\lambda$ becomes the helicity of the one-particle
state. The normalization is such that
\begin{equation}\langle m,j,\vec{p},\lambda
  |m,j,\vec{p}^{\,\prime},\lambda'\rangle =
 (2\pi)^3 2 E(\vec{p}\,) \delta^3(\vec{p}-\vec{p}^{\,\prime})\delta_{\lambda\lambda'}
\end{equation}
with $ E(\vec{p}\,)= \sqrt{m^2+\vec{p}^{\,2}}$.   
Helicity CM two-particle states are defined as
\begin{eqnarray}
|p,\theta,\varphi;\lambda_1,\lambda_2;\gamma\,\rangle=|m_1,j_1;\vec p,\lambda_1\rangle
\otimes \overline{|m_2,j_2;-\vec p,\lambda_2\rangle}\,,\label{eq:twohstates}
\end{eqnarray}
where $\gamma$ encompasses all other not explicitly identified quantum numbers, and

\begin{equation}
\overline{|m_2,j_2;-\vec p,\lambda_2\rangle}=(-1)^{j_2-\lambda_2}R(\varphi,\theta,-\varphi)
R(0,\pi,0) Z_{|\vec{p}\,|}|m_2,j_2;\vec 0,\lambda_2\rangle\,;
\end{equation}
the phase factor $(-1)^{j_2-\lambda_2}$ is introduced so that as
$\vec{p}\to 0$
\begin{equation}
\overline{|m_2,j_2;\vec p=\vec{0},\lambda_2\rangle}= |m_2,j_2;\vec
p=\vec{0},-\lambda_2\rangle 
\end{equation}
Defining the two-particle state in this way guarantees good 
transformation properties under
rotations
\begin{eqnarray}
|p,\theta,\varphi;\lambda_1,\lambda_2; \gamma\,\rangle=R(\varphi,\theta,-\varphi)
|p,0,0;\lambda_1,\lambda_2; \gamma\,\rangle.
\end{eqnarray}
It is convenient to decompose
\begin{eqnarray}
|p,\theta,\varphi;\lambda_1,\lambda_2;\gamma\,\rangle=2\pi \sqrt{\frac{4\sqrt{s}}{|\vec p\,|}}\,
|P\rangle\,|\theta,\varphi;\lambda_1,\lambda_2;\gamma\,\rangle,\label{eq:Hpeque}
\end{eqnarray}
with $P$ the total four-momentum and $P^2=s$. The normalizations
are\footnote{Note that $E_1 E_2\,
  \delta^3(\vec{p}_1-\vec{p}_2)\,\delta^3(\vec{p}^{\,\prime}_1-\vec{p}^{\,\prime}_2) =
  \sqrt{s}\,
  \delta^{4}(P-P')\,\delta^2(\Omega-\Omega')/|\vec{p}\,|$, with
  $\vec{p}=(\vec{p}_1 - \vec{p}_2)/2$.}
\begin{eqnarray}
\langle P'|
  P\rangle&=&(2\pi)^4
\delta^{4}(P-P'),\nonumber\\ 
\langle\theta',\varphi';\lambda'_1,\lambda'_2;\gamma\,'\,
|\theta,\varphi;\lambda_1,\lambda_2;\gamma\,\rangle&=&\delta(\Omega-\Omega')\delta_{\lambda_1\lambda'_1}
\delta_{\lambda_2\lambda'_2}\delta_{\gamma\gamma\,'}.
\end{eqnarray}
The decomposition in Eq.~(\ref{eq:Hpeque}) attends to the fact the
$4$-momentum is a conserved quantity and thus
\begin{equation}
\langle F | S | I \rangle = (2\pi)^4 \delta^{4}(P_F-P_I)\langle \alpha_F | S_P |
\alpha_I \rangle
\end{equation}
and any state of the Hilbert space, containing any number of
particles, can be written as a superposition
of vectors of the form $|P\rangle\,|\alpha\rangle$. The set of
vectors $|\alpha\rangle$ spans the so-called little Hilbert space~\cite{martin1970elementary}. It follows that the scattering operator $S$ may be written as the direct product 
\begin{equation}
S = \identidad\otimes S_P
\end{equation}
such that 
\begin{equation}
\langle F | S | I \rangle = \langle P_F|\identidad|P_I\rangle\langle \alpha_F | S_P | \alpha_I \rangle
\end{equation}
Just as in the case of the $S$ operator, $T$ may also be written as a
direct product
\begin{equation}
T = \identidad\otimes T_P, \qquad S_P = \identidad-iT_P \label{eq:tp}
\end{equation}
This is the form in which the $T$ matrix is generally used. In
fact we refer to $T_P$ as the $T$ operator and
$T_{FI}(s)=\langle \alpha_F | T_P | \alpha_I \rangle$ as the $T-$matrix element.

The CM states can be written in terms of states with well-defined
total angular momentum
\begin{eqnarray}
|p, J,M;\lambda_1,\lambda_2;\gamma\,\rangle=2\pi \sqrt{\frac{4\sqrt{s}}{|\vec p\,|}}\,
|P\rangle\,|J,M;\lambda_1,\lambda_2;\gamma\,\rangle,
\end{eqnarray}
with
\begin{eqnarray}
\langle J',M';\lambda'_1,\lambda'_2;\gamma\,'\,|J,M;\lambda_1,\lambda_2;\gamma\,\rangle=\delta_{JJ'}
\delta_{MM'}\delta_{\lambda_1\lambda'_1}
\delta_{\lambda_2\lambda'_2}\delta_{\gamma\gamma\,'}. \label{eq:ortJM}
\end{eqnarray}
Starting from the case $\theta=0,\varphi=0$
\begin{eqnarray}
|0,0;\lambda_1,\lambda_2;\gamma\,\rangle=\sum_J
C_J\,|J, J_z=\lambda;\lambda_1,\lambda_2;\gamma\,\rangle,\label{eq:whyMisfixed}
\end{eqnarray}
with $\lambda=\lambda_1-\lambda_2$, one arrives at
\begin{eqnarray}
|\theta,\varphi;\lambda_1,\lambda_2;\gamma\,\rangle=\sum_{J,M}
C_J\, {\cal D}^{(J)}_{M\lambda}(\varphi,\theta,-\varphi)\,
|J,M;\lambda_1,\lambda_2;\gamma\,\rangle,\ \  \lambda=\lambda_1-\lambda_2,\label{eq:cJ}
\end{eqnarray}
where ${\cal D}^{(J)}_{MM'}(\alpha,\beta,\gamma)$ is the matrix
representation of a rotation operator $R(\alpha,\beta,\gamma)$ in an
irreducible representation space,
\begin{equation}
{\cal D}^{(J)}_{M'M}(\alpha,\beta,\gamma) = e^{-i\alpha
  M'}d^J_{M'M}(\beta)\,e^{-i\gamma M}, \qquad  d^J_{M'M}(\beta) =
\langle JM' | e^{-i\beta J_y}| J M\rangle 
\end{equation}
From the above equation and using that 
\begin{equation}
\int d\Omega \,{\cal D}^{(J)*}_{M\lambda}(\varphi,\theta,-\varphi){\cal
  D}^{(J')}_{M'\lambda}(\varphi,\theta,-\varphi) =
\frac{4\pi}{2J+1}\delta_{JJ'}\delta_{MM'}  \label{eq:ortD}
\end{equation}
it follows that
\begin{eqnarray}
|J,M;\lambda_1,\lambda_2;\gamma\,\rangle&=&\frac{2J+1}{4\pi\,C_J}\int d\Omega\ {\cal
D}^{(J)*}_{M\lambda}(\varphi,\theta,-\varphi)\,|\theta,\varphi;
\lambda_1,\lambda_2;\gamma\,\rangle\nonumber \\ 
&=& \sqrt{\frac{2J+1}{4\pi}}\int d\Omega\ {\cal
D}^{(J)*}_{M\lambda}(\varphi,\theta,-\varphi)\,|\theta,\varphi;
\lambda_1,\lambda_2;\gamma\,\rangle,\ \  \lambda=\lambda_1-\lambda_2 \,,\label{eq:cambio-base}
\end{eqnarray}
where we have made use of the normalization conditions to 
determine $|C_J|^2=\frac{2J+1}{4\pi}$ and have taken the coefficients
$C_J$ to be real. States with well-defined orbital angular momentum $L$ and spin $S$ can be introduced as
\begin{eqnarray}
|J,M;\lambda_1,\lambda_2;\gamma\,\rangle&=&\sum_{L,S}\sqrt{\frac{2L+1}{2J+1}}(L,S,J|0,\lambda,\lambda)
(j_1,j_2,S|\lambda_1,-\lambda_2,\lambda)|J,M;L,S;\gamma\,\rangle,\nonumber\\
|J,M;L,S;\gamma\,\rangle&=&\sum_{\lambda_1,\lambda_2}\sqrt{\frac{2L+1}{2J+1}}(L,S,J|0,\lambda,\lambda)
(j_1,j_2,S|\lambda_1,-\lambda_2,\lambda)|J,M;\lambda_1,\lambda_2;\gamma\,\rangle,\label{eq:LSstates}
\end{eqnarray}
where $(j_1,j_2,j|m_1,m_2,M)$ are the Clebsch-Gordan coefficients and
$\lambda=\lambda_1-\lambda_2$ as usual.

\section{Properties of the $\chi_{r,\lambda}$ amplitudes defined in
  Eq.~(\ref{eq:fase33})}
\label{app:chi}

The $\chi_{r,\lambda}$ amplitudes in Eq.~(\ref{eq:fase33}) can be
rewritten in terms of states $|J,M; L,S\rangle$ with well-defined
total orbital ($L$) and spin ($S$)
angular momenta as
\begin{eqnarray}
\chi_{r,\lambda}&=&\sqrt{\frac{3}{8}} \sum_{\rho}
(1,1/2,3/2|0,-\rho,-\rho)\,\langle
3/2,M;{0,\rho}|T(s)|3/2,M;{r,\lambda}\rangle\nonumber
\\
&=&\frac{1}{\sqrt{2}}\langle
3/2,M;{L=1,S=1/2}|T(s)|3/2,M;{r,\lambda}\rangle\nonumber\\
&=&\sum_{L',S'}\sqrt{\frac{2L'+1}{8}}(1,1/2,S'|r,-\lambda,M)(L',S',3/2|0,M,M)
\langle {L=1,S=1/2}|T_{J=\frac32}(s)|{L'S'}\rangle,
\label{eq:nodepM}
\end{eqnarray}
with $M=r-\lambda$. Note that the matrix element of the $T$ scattering operator does not depend on $M$,
since it is invariant under rotations.  Now,  the amplitude has a vector 
and an axial part,
\begin{eqnarray}
{ T}={ T}^{ V}-{ T}^{ A}, 
\end{eqnarray}
and under a parity transformation, we have
\begin{eqnarray}
{\cal P}\,{ T}\,{\cal P}^\dagger&=&{ T}^{ V}+ T^{ A} \\
{\cal P}|J,M; L S\rangle&=&\eta_1\eta_2(-1)^{L}|J,M; L,S\rangle
\end{eqnarray}
where $\eta_{1,2}$ are the intrinsic parities of the particles (1 for
nucleons and $-1$ for  $\pi$ and $W$). We thus find that only odd
(even) $L'$ waves contribute to the vector (axial) part of the $\chi_{r,\lambda}$,
\begin{eqnarray}
\chi_{r,\lambda}&=&\chi_{r,\lambda}^V -
 \chi_{r,\lambda}^A \\
\chi_{r,\lambda}^V &=&\sum_{S'=1/2,3/2}\sum_{L'=1,3}\sqrt{\frac{2L'+1}{8}}(1,1/2,S'|r,-\lambda,M)(L',S',3/2|0,M,M)
\langle {L=1,S=1/2}|T^V_{J=\frac32}(s)|{L'S'}\rangle\\
\chi_{r,\lambda}^A &=&\sum_{S'=1/2,3/2}\sum_{L'=0,2}\sqrt{\frac{2L'+1}{8}}(1,1/2,S'|r,-\lambda,M)(L',S',3/2|0,M,M)
\langle {L=1,S=1/2}|T^A_{J=\frac32}(s)|{L'S'}\rangle
\end{eqnarray}
Now taking into account 
\begin{equation}
(1,1/2,S'|r,-\lambda,M)(L',S',3/2|0,M,M) = (-1)^{L'}(1,1/2,S'|-r,\lambda,-M)(L',S',3/2|0,-M,-M)
\end{equation}
we trivially find Eq.~(\ref{eq:relPJ}).

On the other hand, using the basis introduced in
Eq.~(\ref{eq:baseTD}), we obtain the following relations\footnote{For matrix elements of the vector current, the involved multipoles in the notation of Ref.~\cite{Drechsel:1992pn} are shown in square brackets.}
\begin{eqnarray}
-\frac12\left\langle L=1,S=1/2\,| T^V_{J=3/2}|L'=1,\tilde l=1\right\rangle &=&
\frac12\left(\chi^V_{1,1/2}+\sqrt{3}\chi^V_{1,-1/2}
\right),\, [M_{1+}]\label{eq:dominantv}\\
-\frac12\left\langle L=1,S=1/2\,| T^V_{J=3/2}|L'=1,\tilde l=2\right\rangle &=&
\frac{1}{\sqrt{20}}\left(2\sqrt{2}\chi^V_{0,-1/2}+\sqrt{3}\chi^V_{1,-1/2}-3\chi^V_{1,1/2}
\right),\, [E_{1+}/L_{1+}] \\
-\frac12\left\langle L=1,S=1/2\,| T^V_{J=3/2}|L'=3,\tilde l=2\right\rangle &=&
\frac{1}{\sqrt{10}}\left(-\sqrt{6}\chi^V_{0,-1/2}+\chi^V_{1,-1/2}-\sqrt{3}\chi^V_{1,1/2}
\right),\, [E_{1+}/L_{1+}] \\
-\frac12\left\langle L=1,S=1/2\,| T^A_{J=3/2}|L'=0,\tilde l=1\right\rangle &=&
-\frac{1}{\sqrt{6}}\left(\sqrt{2}\chi^A_{0,-1/2}+\sqrt{3}\chi^A_{1,-1/2}+\chi^A_{1,1/2}
\right) \label{eq:dominanta}\\
-\frac12\left\langle L=1,S=1/2\,| T^A_{J=3/2}|L'=2,\tilde l=1\right\rangle &=&
-\frac{1}{\sqrt{12}}\left(-2\sqrt{2}\chi^A_{0,-1/2}+\sqrt{3}\chi^A_{1,-1/2}+\chi^A_{1,1/2}
\right)\label{eq:subdominanta}\\
-\frac12\left\langle L=1,S=1/2\,| T^A_{J=3/2}|L'=2,\tilde l=2\right\rangle &=&
-\frac12\left(\chi^A_{1,-1/2}-\sqrt{3}\chi^A_{1,1/2}
\right)\label{eq:dominantfinal}
\end{eqnarray}
\section{Computation of the $\chi_{r,\lambda}(s)$
amplitudes within the HNV model}
\label{app:chiHNV}
Equation~(\ref{eq:watson-part}) allows us to compute $\chi_{r,\lambda}(s)$ in terms of the matrix elements
$\underbrace{\langle \theta,\varphi;{0,\rho}}_{\pi^+p}|T(s)|\underbrace{0,0;{r,\lambda}\rangle}_{W^+p}$, 
which involve the helicity CM two-particle states introduced in Eq.~(\ref{eq:twohstates}).
We have always labeled the proton as the second particle.  This
is to say that the ``bar'' $\overline{|j; -\vec p,\lambda\rangle}$ states correspond to the protons.
One can prove that 
\begin{equation}
\overline{|j;-\vec
  p,\lambda\rangle}=(-1)^{j-\lambda}(-1)^{2j}e^{-2i\lambda\varphi}|j;-\vec{p},\lambda\rangle  \label{eq:umenospbarra}
\end{equation}
with 
\begin{equation}
|j;-\vec{p},\lambda\rangle = R(\varphi+\pi,\pi-\theta,-\varphi-\pi)
Z_{|\vec{p}\,|}|j;\vec 0,\lambda\rangle, \label{eq:umenosp}
\end{equation}
where $\theta$ and $\varphi$ are the polar and azimuthal angles of
$\vec{p}$. The latter states, $|j;-\vec{p},\lambda\rangle$ for the case of a
nucleon $j=1/2$, can be easily obtained using the Dirac space representations 
of the boost and the rotation  that appears in
Eq.~(\ref{eq:umenosp}). Finally, and using  
Eq.~(\ref{eq:umenospbarra}), we find that the spinors corresponding to the bar states are 
\begin{eqnarray}
\overline{\left.\Big|-\vec{p},\lambda=1/2\right\rangle} &\equiv& \sqrt{E+M_N}\left ( \begin{array}{c}-\sin\frac{\theta}{2}
e^{-i\varphi}\\
\cos\frac{\theta}{2}\\-\frac{|\vec{p}\,|}{E+M_N}\sin\frac{\theta}{2} e^{-i\varphi}\\
\frac{|\vec{p}\,|}{E+M_N}\cos\frac{\theta}{2}\end{array}\right) \label{eq:bar1}\\
\nonumber\\
\overline{\left.\Big|-\vec{p},\lambda=-1/2\right\rangle} &\equiv& \sqrt{E+M_N}\left ( \begin{array}{c}\cos\frac{\theta}{2}\\
\sin\frac{\theta}{2}e^{i\varphi}\\-\frac{|\vec{p}\,|}{E+M_N}
\cos\frac{\theta}{2}\\
-\frac{|\vec{p}\,|}{E+M_N}\sin\frac{\theta}{2}e^{i\varphi}\end{array}\right) \label{eq:bar2}
\end{eqnarray}
with $M_N$ the nucleon mass and $E=\sqrt{M_N^2+\vec{p}^{\,2}}$. On the
other hand, the virtual gauge boson helicity states, when the $W$ three-momentum
 is in the positive $z$ direction, read
\begin{eqnarray}
\epsilon^\mu(|\vec{p}\,|, r=0)&=&
\left(|\vec{p}\,|/\sqrt{Q^2},0,0,\sqrt{\vec{p}^{\,2}-Q^2}/\sqrt{Q^2} \right) \label{eq:pol_r0}\\
\epsilon^\mu(|\vec{p}\,|, r=\pm 1)&=& \frac{\mp1}{\sqrt2} \left(0,1,\pm i, 0 \right)
\end{eqnarray}
with $-Q^2$,
the virtual mass of the gauge boson, i.e., its four-momentum
squared. We only consider the three polarizations that are orthogonal to the
$W$ four-momentum since our analysis in Secs.~\ref{sec:wtwz} and \ref{sec:olss}
implicitly assumes a positive invariant mass squared for the $W$  boson. The
results are then analytically continued to
  negative invariant masses squared.

With all of these ingredients, 
within the HNV model we deduce that, up to an overall real 
normalization constant that does not affect
its phase, 
\begin{equation}
\langle
  \theta,\varphi;{0,\rho}|T(s)|0,0;{r,\lambda}\rangle \sim -i\, \left[j_\mu(\rho,\lambda)
\epsilon^\mu(|\vec{p}\,|, r)\right], \label{eq:eq3glup}
\end{equation}
where the $p\pi^+$ current  $j^\mu$ is taken from Eq.(51)
of Ref.~\cite{Hernandez:2007qq} and Eq.~(A6) of Ref.~\cite{Hernandez:2013jka}, replacing the proton spinors
$u(\vec{p}\,)$ by the ``bar'' states of Eqs.~(\ref{eq:bar1}) and
(\ref{eq:bar2}) corresponding\footnote{Obviously, the $\bar{u}$
  spinor that appears in Eq.(51) of Ref.~\cite{Hernandez:2007qq}
  should be evaluated using Eqs.~(\ref{eq:bar1}) and (\ref{eq:bar2}), 
  taking Hermitian conjugation ($\dagger$) and multiplying by the $\gamma^0$ Dirac matrix.} to the helicities $\rho$ and
$\lambda$. The current of Eq.~(A6) of Ref.~\cite{Hernandez:2013jka}
accounts for the crossed $N(1520)$ pole mechanism, which gives a
quite small contribution for the $\pi N$ invariant masses studied in
this work. Note 
that the direct $N^*(1520)$ excitation mechanism also considered in
Ref.~\cite{Hernandez:2013jka} does not contribute to the isospin 3/2 channel. 

Note that in the definition of the current $j^\mu$ in
Refs.~\cite{Hernandez:2007qq,Hernandez:2013jka}, the factor $i$ from
the weak vertex is not included. Actually the gauge coupling is not
included either. According to our normalizations, one has
$-iT_{\rm  aux} = i {\cal L } \propto i\, j^\mu \epsilon_\mu$ and thus,
up to real constants, $T_{\rm aux}$ is given by
$-j\cdot\epsilon$. The extra $i$ ($i T_{\rm aux} = T)$ in
Eq.~(\ref{eq:eq3glup})  is included to
ensure that Eq.~(\ref{eq:2glup}) that leads to Eqs.~(\ref{eq:1glup})
and (\ref{eq:4glup})  is
satisfied. 
This is needed because in our conventions the pion and
the $W$ gauge boson intrinsic time reversal phases are different ($-1$ and
1, respectively). To keep Eq.~(\ref{eq:2glup}) correct, one
should add a phase $i$ to the $\pi N$ state, which compensates the pion
odd intrinsic time reversal\footnote{This is easy to see for instance
  by looking at the $\pi NN$ Lagrangian in Eq.~(26) of
  Ref.~\cite{Hernandez:2007qq}, and considering the transformation under
  time reversal of the nucleon axial current and the derivative
  operator. } thanks to the antiunitary character of the
time-reversal operator in Eq.~(\ref{eq:2glup}).

In addition and to implement Watson's theorem, within the
approximate Olsson scheme discussed in Sec.~\ref{sec:olss}, the
vector and axial  direct $\Delta$ contributions should be multiplied
by  the Olsson phases, $\Psi_V$ and $\Psi_A$.

\section{parametrizations of the $\Psi_V$ and $\Psi_A$ Olsson phases}
\label{app:fits}
In the following, we give  parametrizations 
for the $\Psi_V$ and $\Psi_A$ Olsson phases, as a function of 
$w=W_{\pi N}-1.0779\,$GeV and $Q^2=-q^2$, 
valid in the intervals $W_{\pi N}\in[1.1,1.4]\,$GeV,  $Q^2\in
[0,2.5]\,$GeV$^2$. 
\begin{enumerate}
\item Fit A:
\begin{eqnarray}
\Psi_V=5w\,\bigg(8.3787 + \frac{2.7315 - 25.5185\, w}{
     	  0.05308416 + (0.62862 - 5\,w)^2} + 301.925\,w - 
     	 985.80\,w^2 + 862.025\,w^3\bigg)\nonumber\\
       \times \bigg((1. + 0.14163\,Q^2)^{-2} + 
     	 \big(0.066192 + w\,(-0.34057 + 1.631475\,w)\,\big)\,Q^2\bigg), 
\end{eqnarray}
\begin{eqnarray}
\Psi_A=5w\,\bigg(5.2514 + \frac{2.9102 - 26.5085\, w}{
     	  0.0531901969 + (0.63033 - 5\,w)^2} + 266.565\,w - 
     	 814.575\,w^2 + 624.05\,w^3\bigg)\nonumber\\
       \times \bigg((1. + 0.088539\,Q^2)^{-2} + 
     	 \big(0.026654 + w\,(-1.17305 + 3.66475\,w)\,\big)\,Q^2\bigg). 
\end{eqnarray}
\item Fit B:
\begin{eqnarray}
\Psi_A=5w\,\bigg(4.9703 + \frac{2.929 - 26.6295\, w}{
     	  0.0531256401 + (0.63051 - 5\,w)^2} + 264.27\,w - 
     	 798.525\,w^2 + 598.85\,w^3\bigg)\nonumber\\
       \times \bigg((1. + 0.10152\,Q^2)^{-2} + 
     	 \big(0.041484 + w\,(-1.20715 + 3.7545\,w)\,\big)\,Q^2\bigg), 
\end{eqnarray}
while $\Psi_V$ is the same as for fit A.
\end{enumerate}

\bibliography{neutrinos}

\end{document}